\documentclass[10pt,a4paper]{article}
\usepackage{geometry}
\usepackage[utf8]{inputenc}

\usepackage{caption}
\usepackage{subcaption}
\usepackage{jcappub}
\usepackage{bm}
\usepackage{epsfig}
\usepackage{bbold}
\usepackage{graphicx}
\usepackage{amsmath}
\usepackage{amssymb}
\usepackage{amsbsy}
\usepackage{color}
\usepackage{slashed}
\usepackage{afterpage}
\usepackage{psfrag}
\usepackage[noabbrev]{cleveref}
\usepackage{dsfont}
\usepackage{enumerate}
\usepackage[shortlabels]{enumitem}
\usepackage[utf8]{inputenc}
\usepackage{amsmath}
\usepackage[dvipsnames]{xcolor}
\usepackage{graphicx}
\usepackage{braket}

\def\subt#1#2{#1_\text{#2}}

\title{\begin{center}
CMB and Lyman-$\alpha$ constraints on dark matter decays to photons
\end{center}}

\author[a]{Francesco Capozzi,}
\author[b]{Ricardo Z. Ferreira,}
\author[c]{Laura Lopez-Honorez}
\author{and}
\author[d]{Olga Mena}

\affiliation[a]{Dipartimento di Scienze Fisiche e Chimiche, Universita degli Studi dell’Aquila, 67100 L’Aquila,
Italy}
\affiliation[b]{Institut de Física d’Altes Energies (IFAE) and Barcelona Institute of Science and Technology (BIST), Campus UAB, 08193 Bellaterra, Barcelona, Spain}
\affiliation[c]{Service de Physique Theorique, Universite Libre de Bruxelles, C.P. 225, B-1050 Brussels, Belgium.\\
Theoretische Natuurkunde \& The International Solvay Institutes,
Vrije Universiteit Brussel, Pleinlaan 2, B-1050 Brussels, Belgium}
\affiliation[d]{Instituto de F\'{i}sica Corpuscular (IFIC), University of Valencia-CSIC, Parc Cient\'{i}fic UV, c/ Cate\-dr\'{a}tico Jos\'{e} Beltr\'{a}n 2, E-46980 Paterna, Spain}

\emailAdd{francesco.capozzi@univaq.it}
\emailAdd{rzambujal@ifae.es}
\emailAdd{Laura.Lopez.Honorez@ulb.be}
\emailAdd{omena@ific.uv.es}

\makeatletter
\gdef\@fpheader{}
\makeatother

\abstract{
Dark matter energy injection in the early universe modifies both the ionization history and the temperature of the intergalactic medium. 
In this work, we improve the CMB bounds on sub-keV dark matter and extend previous bounds from Lyman-$\alpha$ observations to the same mass range, resulting in new and competitive constraints on axion-like particles (ALPs) decaying into two photons. 
The limits  depend on the underlying reionization history, here accounted self-consistently by our modified version of the publicly available {\tt DarkHistory}  and {\tt CLASS} codes. Future measurements such as the ones from the CMB-S4 experiment
may play a crucial, leading role in the search for this type of light dark matter candidates.}

\begin{document}
\hfill{\small ULB-TH/23-03}
\maketitle

\vspace{0.3cm}

\section{Introduction}
\label{sec:intro}

Planck observations of the Cosmic Microwave Background Anisotropies (CMB)~\cite{Planck:2018vyg} constrain the cosmic ionization history while Lyman-$\alpha$ data, combined with state-of-the-art hydrodynamical simulations, have allowed precise determinations of the  intergalactic medium (IGM) temperature at low-redshifts~ (see e.g. \cite{Walther:2018pnn,Gaikwad:2020art} for recent analysis). These observations have  then recently been used to search for the effects of annihilations and decays of Dark Matter  (DM) particles that are known to modify both the ionization history and the temperature of the IGM throughout the universe's history.

In this work, we revisit the imprints on the ionization history, from the recombination period until present times, and on the IGM temperature, at low redshifts ($z\lesssim 6$), for DM decays into two photons. We exploit Planck 2018 data to update previous CMB constraints on the 20.4 eV to keV mass range, and Lyman-$\alpha$ data to extend previous analyses for heavier DM masses to the same mass window, in which a plethora of axion-like particle (ALPs) DM models may lie.~\footnote{See e.g. Refs.~\cite{Takahashi:2021tff, Gelmini:2021yzu, Bernal:2022xyi, Branco:2023frw,Carenza:2023qxh} for recent studies of ALP DM in this mass range.} 
The lower end of the mass range corresponds to twice the energy necessary for a Lyman-$\alpha$ transition in the Hydrogen atom: a photon with an energy below the Lyman-$\alpha$ threshold interacts with the gas much more weakly than a photon above the threshold, see e.g.~\cite{Slatyer:2015jla,Slatyer:2015kla}. At the upper end, strong constraints from X-ray searches~\cite{Cadamuro:2011fd} dramatically prevent us to improve over the existing bounds with CMB and Lyman-$\alpha$ data.

For the CMB constraints, we will consider Planck 2018 data and extend the work of \cite{Cadamuro:2011fd,Bolliet:2020ofj} in a few ways.  
First, we take into account the energy injection efficiencies by making use of the {\tt DarkHistory} code~\cite{Liu:2019bbm} and investigating the impact on the bounds of multiple reionization scenarios. We consider two well-motivated astrophysical models for the galactic UV/X-ray background~\cite{Faucher-Giguere:2019kbp,Puchwein:2018arm} and self-consistently take into account the DM feedback on the IGM temperature and on the ionization fractions by means of the use of {\tt DarkHistory}. We then perform a full MCMC analysis in which we vary not only the relevant DM  parameters but also other fiducial cosmological parameters, which can exhibit degeneracies, thus deriving robust bounds on both the DM mass and its coupling to photons. 
We find bounds that can be competitive with those from the Leo-T dwarf galaxy \cite{Wadekar:2021qae}~\footnote{
Gas-rich dwarf galaxies exhibit a behavior close to primitive galaxies in the early Universe and therefore they can be exploited as a tool to constrain non-standard cosmic ionization histories \cite{Wadekar:2021qae}.}, when the astrophysical reionization model yields a relatively  large optical depth to reionization. On the other hand,  we show that CMB bounds are expected to become competitive to those of Leo-T  with future CMB surveys, independently of the assumed reionization history.

Concerning the Lyman-$\alpha$ data analysis, we shall derive new bounds by extending the analysis provided in~\cite{Liu:2020wqz} to lower DM masses. We will follow the conservative approach proposed in Ref.~\cite{Liu:2020wqz}, where robust constraints on DM from the IGM temperature were derived by fixing the reionization history to the Planck fiducial model and by neglecting the photoheating from astrophysical sources thus overcoming the large uncertainties associated to the astrophysical scenarios.

The structure of the manuscript is as follows. Section~\ref{sec:sec2} contains a short description of the evolution of both the IGM temperature  $T_m$ and the free electron fraction $x_e$, including different reionization models.  
In Sec.~\ref{sec: CMB analysis}, we introduce the treatment of the energy deposition efficiency that we then employ in our up-to-date CMB 
analysis to derive constraints on light DM decaying to photons with current data. In the same section, we also forecast how future CMB experiments will improve over the current constraints.
 Section~\ref{sec:Lycons} describes the analysis with Lyman-$\alpha$ data and the resulting constraints on the same DM decaying to photons scenario. Finally, we draw our conclusions in Sec.~\ref{sec:concl}.

\section{Ionized fraction and IGM temperature evolution}
\label{sec:sec2}
We start by briefly reviewing the different contributions to the evolution of the temperature of the intergalactic medium (IGM) $T_m$, and of the free electron fraction $x_e$.
Here, we follow closely Refs.~\cite{Liu:2019bbm,Liu:2020wqz},  whose formalism has been implemented in the publicly available code {\tt DarkHistory}~\cite{Liu:2019bbm}. This code allows to systematically solve for $T_m$ and $x_e$ including DM injections of energy all along recombination and reionization history as well as specific astrophysical models for the photoionization and photoheating rates at low redshifts. We will make use of {\tt DarkHistory} (with some modifications) to obtain the results presented in Secs~\ref{sec: CMB analysis} and \ref{sec:Lycons}.

The evolution of the different ionization fractions is entangled with the evolution of the IGM temperature.  The system of equations that keeps track of $T_m$ and of the different contributions to $x_e$ reads~\cite{Peebles:1968ja,Zeldovich:1969ff}:
\begin{eqnarray}
    \label{eq: master equation for Tm and x}
 \dot{Y} = \dot{Y}^{(0)} + \dot{Y}^\text{DM} + \dot{Y}^\text{astro}, \qquad {\rm where}\quad Y= \left(\begin{matrix}  T_m \\ x_\text{HII} \\ x_\text{HeII} 
 \\ x_\text{HeIII}  \end{matrix} \right) \,,
\label{eq:dotY}
\end{eqnarray}
where the ionized fractions $x_X$ correspond to the ratios $x_X= \subt{n}{X}/n_\text{H}$ where $n_H$ is the total Hydrogen density and  $X=$ HII, HeII and HeIII stands for Hydrogen, singly ionized Helium and doubly ionized Helium, respectively.~\footnote{Notice that  at high redshifts the  ionized  Helium contributions can be neglected and the free electron fraction $x_e$ reduces to the Hydrogen ionized fraction $x_\text{HII}$.}  The contributions to the evolution of the temperature and ionized fractions are divided into three different terms. The first term $\dot Y^{(0)}$  accounts for  adiabatic evolution, Compton scatterings and atomic processes, while the $\dot Y^{\rm DM}$ term  is driven by DM  energy injection in the medium. The third term $\dot Y^{\rm astro}$ is particularly relevant at low redshifts when astrophysics sources provide an extra source of  photoionization and photoheating,  triggering reionization. In Secs.~\ref{sec:adiab} to \ref{sec:reio} below, we briefly discuss each of these terms.

\subsection{Adiabatic cooling, Compton scattering and atomic processes}
\label{sec:adiab}

Let us start by describing the $\dot Y^{(0)}$ term in Eq.~(\ref{eq:dotY}). The corresponding contribution to the IGM temperature evolution reads \cite{Liu:2019bbm,Liu:2020wqz}:
\begin{eqnarray}
	\dot{T}_m^{(0)} =  -2 H T_m+ \Gamma_C\left(\subt{T}{CMB}-T_m \right) + \dot{T}_m^\text{atom} \, .
\end{eqnarray}
The first term accounts for adiabatic cooling whereas the second term describes Compton heating/cooling with  $\Gamma_C$ the Compton scattering rate, $\subt{T}{CMB}$  the CMB temperature, and $H$ is the Hubble rate. The last term includes multiple heating/cooling contributions due to atomic processes (recombination, collisional ionization, collisional excitation and bremsstrahlung) whose rates are given in  Refs.~\cite{Theuns:1998kr,Bolton:2006pc} (see also \cite{Liu:2020wqz}). 
On the other hand, the evolutions of the ionized fractions is governed by 
\begin{eqnarray}
	\label{eq: adiabatic terms, ionization fractions}
 \dot{x}_\text{X}^{(0)} = 	 \dot{x}_\text{X}^\text{ion} -\dot{x}_\text{X}^\text{rec}\, ,
\end{eqnarray}
where $x_X^{\rm ion}$ ($x_X^{\rm rec}$) accounts for ionization (recombination) processes. 
It is customary to discriminate between situations in which recombinations to the ground state are accounted for (case-A) or not (case-B), see e.g.~\cite{Bolton:2006pc,Liu:2019bbm}. In our analysis, we follow~\cite{Liu:2019bbm} that treats redshifts below and above the onset of reionization ($z_A^\text{max}$) differently. At large redshifts compared to $z_A^\text{max}$, the universe is optically thin so case-B recombination and photoionization coefficients apply and the $\dot x_ X^{(0)}$ term includes both contributions.\footnote{In an optically thin medium, photons of 13.6 eV, arising from recombinations to the $H$ ground state, are absorbed in much less than a Hubble time and Hydrogen cannot recombine in this way. In the latter case, only case-B recombination coefficients shall be taken into account.} In contrast, for $z\lesssim z_A^\text{max}$, the universe is more opaque to light and case-A  recombination and  collisional ionization contributions are instead taken into account (see \cite{Theuns:1998kr,Bolton:2006pc,Liu:2019bbm} and Appendix \ref{sec: Rates} for the rates). 
Note that, at low redshifts, the photoionization rates are dominated by astrophysical contributions that are included  in the $\dot{Y}^\text{astro}$ term of Eq.~(\ref{eq: master equation for Tm and x}) (see the discussion in Sec.~\ref{sec:astro}). 

Finally, we briefly comment on a modification in the matter temperature evolution considered in our analysis compared to the default implementation in  {\tt DarkHistory}. In the latter code,
collisional excitation processes
are only included at low redshift, after recombination starts. We found, however, that this is not a good approximation for light dark matter masses below ${\cal O}(100)$ eV (see Appendix \ref{sec:modif_darkhistory} and Fig. \ref{fig:tm_xe_modified_darkhistory}). 
Therefore, we have modified the code to include the effect of collisional excitation at all times for the matter temperature evolution.

\subsection{Dark matter energy  injection and deposition }
\label{sec:DM}

The $\dot Y^{\rm DM}$ term  of Eq.~(\ref{eq: master equation for Tm and x}) accounts for the  dark matter annihilation/decay contributions. To describe this term, let us  focus on an ALP dark matter particle $a$ of mass $m_a$ that decays into two photons of energy $m_a/2$ at a rate $\Gamma_\text{dec}\gg t_0^{-1}$  (where $t_0$ is the age of the universe). The energy injected per unit of time and volume is given by
\begin{equation}
	\left(\frac{dE (z)}{dt\,dV}\right)_{\rm injected}=\rho_a(1+z)^3\Gamma_\text{dec}\,,
	\label{eq:inj}
\end{equation}
where $\rho_a$ is the energy density of the DM particle today and the decay rate is parametrized as 
\begin{eqnarray} \label{eq: Axion decay rate}
    \Gamma_\text{dec} =g_{a\gamma \gamma}^2 m_a^3/(64\pi)~,
\end{eqnarray}
with  $g_{a \gamma \gamma}$ the ALP-photon coupling. In the next sections, we will phrase our constraints on DM decays to photons in terms of the ALPs parameters $m_a$ and $g_{a\gamma \gamma}$. Note  however that, by properly re-expressing the bounds on $g_{a\gamma \gamma}$ in terms of the DM lifetime $\Gamma_\text{dec}^{-1}$, our constraints  apply to any DM model decaying to two photons.

The injected energy may not be deposited instantaneously into the medium due to the cooling of primary particles. In addition, there are multiple channels $c$ of energy deposition  including IGM heating (denoted with $c=$ heat),  Hydrogen ionization ($c=$ HII),    Helium single or double ionization ($c=$ HeII or HeIII), and neutral atom excitation ($c=$ exc).
The fraction of energy injected that is deposited in the different channels can be expressed as~\cite{Slatyer:2009yq}
\begin{equation}
	\left(\frac{dE_c (x_e,z)}{dt\,dV}\right)_{\rm deposited} =   f_c(x_e,z) \, \left(\frac{dE (z)}{dt\,dV}\right)_{\rm injected}~,
	\label{eq:dep}
\end{equation}
where the coefficients,  $f_c(x_e, z)$,  are the DM energy deposition efficiencies. 
 They account for  all the details associated to the delay in energy deposition and separation into different channels $c$ at a given redshift $z$ and free electron fraction $x_e$ (that is a function of the different ionization fractions $x_{\rm X}$).\footnote{In practice, $f_c$ depends on each of the ionization fraction $x_\text{H}, x_\text{HeII}, x_\text{HeIII}$ independently~\cite{Liu:2019bbm}.} 
We make use of {\tt DarkHistory}~\cite{Liu:2019bbm} to obtain the $f_c(x_e,z)$ functions. In the {\tt DarkHistory} code, the term $\dot{Y}^{\text{DM}}$ in Eq.~(\ref{eq: master equation for Tm and x}) takes the form:
\begin{eqnarray} \label{eq: Contributions from DM injection}
\dot{Y}^{\text{DM}} = A  \times \frac{1}{\subt{n}{H}} 	\left(\frac{dE (z)}{dt\,dV}\right)_{\rm injected} ~,
\end{eqnarray}
where the prefactor $A=A(f_c(x_e,z))$ is a function of the deposition fractions $f_c(x_e,z)$ (see Appendix~\ref{sec: Rates} and  Ref.~\cite{Liu:2020wqz} for details).

\subsection{Reionization}
\label{sec:reio}

\begin{figure}
  \centering
    \includegraphics[width=0.46\textwidth]{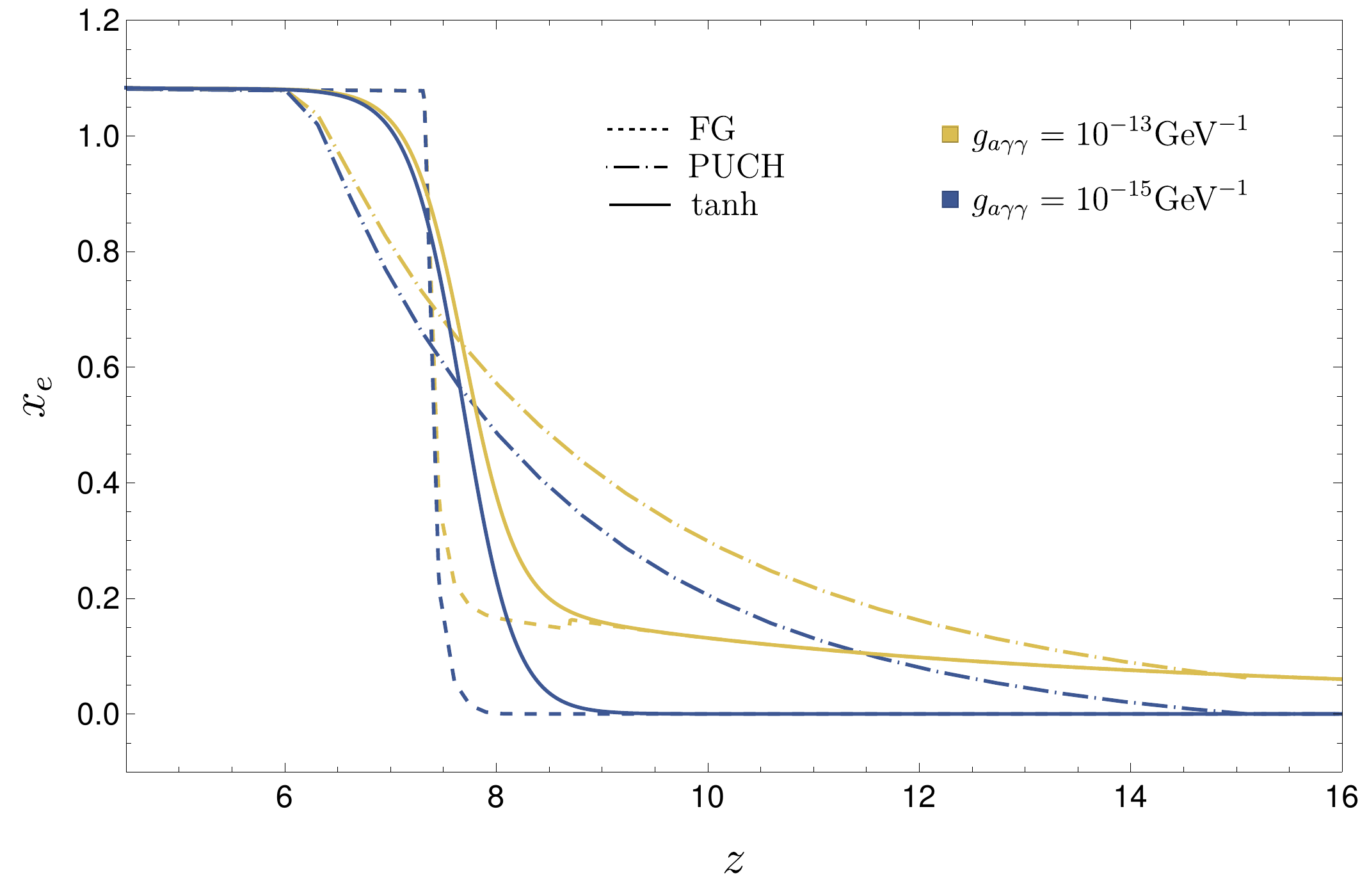}
    \includegraphics[width=0.46\textwidth]{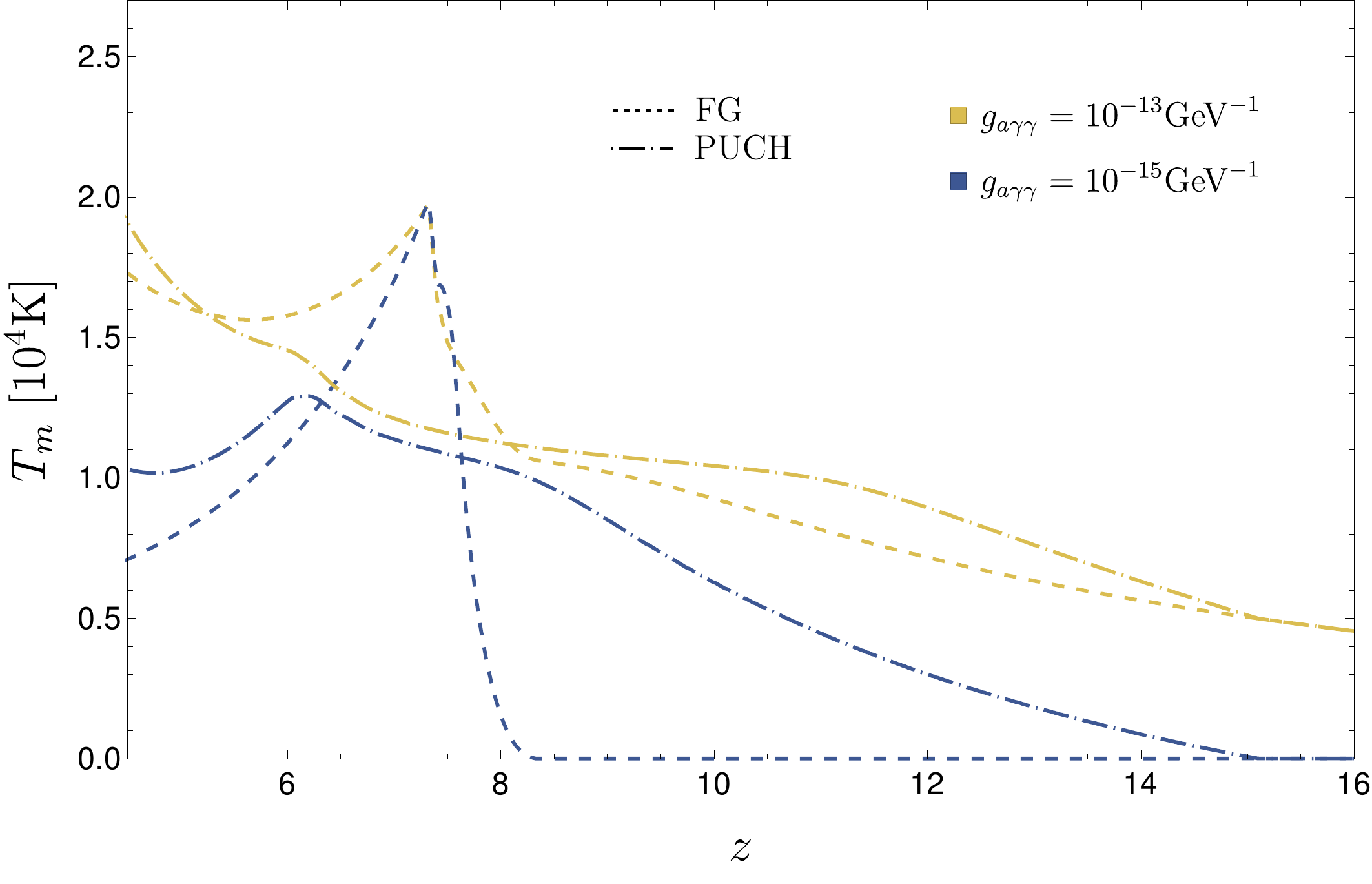} \caption{\textit{\textbf{Ionisation and IGM temperature histories }} for different reionization models including the energy injections 
 from the decays into two photons of a DM particle of mass $m_a=~95$ eV and two different couplings to photons. On the left panel, we focus on the free electron fraction illustrating with blue and orange curves the reionization models of FG (dashed lines), PUCH (dot-dashed lines) and hyperbolic tangent (solid lines) for $z_{\rm reio} =7.68$ (continuous) and consider DM energy injection for two possible values of the DM-photon coupling $g_{a\gamma\gamma}:  10^{-15}$ and $10^{-13}$~GeV$^{-1}$. The right panel depicts the matter temperature evolution in redshift for the very same two astrophysical-based reionization models and couplings  $g_{a\gamma\gamma}$.}
\label{fig:comparison_PUCH_FG}
\end{figure}

At low redshifts, typically at $z \lesssim z_A^\text{max}$, star formation and active galactic nuclei are expected to inject extra sources of energy in the IGM and to drive reionization at $z=z_{\rm reio}<z_A^\text{max}$, see e.g. the discussion in Refs.~\cite{Lopez-Honorez:2013cua, Poulin:2015pna}.
Here, we will consider two different approaches to reionization. 
We begin with a description of the canonical hyperbolic tangent model in Sec.~\ref{sec:tanh}. The latter provides an 
effective parametrization of the ionized fraction at low $z$. 
We then describe in Sec.~\ref{sec:astro} two well-motivated
reionization scenarios that rely on different models of UV and
X-ray background emission  from galaxy formation processes  and give rise to distinct photoionization and photoheating rates.
The latter are  necessary to make  a self-contained evolution of both the free electron fraction, $x_e$,
and the IGM temperature, $T_m$, at low $z$. Their effect is taken into account in the $\dot Y^{\rm astro}$ term of Eq.~(\ref{eq:dotY}). The corresponding ionized fractions and, when relevant, temperature evolutions, are
illustrated in Fig.~\ref{fig:comparison_PUCH_FG} around the epoch of reionization for each of the models.
In our CMB analysis of Sec.~\ref{sec: CMB analysis}, we  study the impact of the ionized fraction $x_e$ evolution, obtained for the three different reionization models, on the CMB anisotropies.
In Sec.~\ref{sec:Lycons}, when computing the Lyman-$\alpha$ constraints on $T_m$, we will instead be interested in deriving conservative bounds on the amount of DM heating thus, following~\cite{Liu:2020wqz}, we neglect the astrophysical sources of
heating at low redshifts and fix the ionized fraction at low redshifts to the hyperbolic tangent model.

A useful quantity when comparing different reionization histories is
the optical depth to reionization, $\tau$, that we define as
\begin{equation}
    \tau=\int^{z_{e, \rm min}}_0 dz\, n_e \sigma_T\, \frac{dt}{dz}\,,
    \label{eq:tau}
\end{equation}
i.e. the integral, 
between today and the time at which the electron fraction displays a
minimum (tagged as $z_{e, \rm  min}$),   of the free electron
number density\footnote{ $n_e(z)=x_e(z) n_{\rm{H}}(z)$, where $n_{\rm{H}}$ is the Hydrogen 
density.}, $n_e (z)$, multiplied by the Thompson cross-section, $\sigma_T$.
This is the prescription followed in the publicly available  {\tt
  CLASS} Boltzmann solver code~\cite{Lesgourgues:2011re,Blas:2011rf,Lesgourgues:2013bra}.\footnote{Other prescriptions might have defined a default maximum redshift. In the case of e.g. extended reionization histories or exotic energy injection, an arbitrary choice of maximum redshift  strongly affect the value of $\tau$ while the prescription used in Eq.~(\ref{eq:tau}) is nearer to what Planck data is effectively sensitive to, see the discussion in Ref.~\cite{Poulin:2015pna}.}

\subsubsection{The hyperbolic tangent function}
\label{sec:tanh}

The most widely used model for the reionization history exploits the
hyperbolic tangent function~\cite{Lewis:2008wr}:
\begin{equation} 
x_e^{\rm tanh}(z) = \frac{1+{\cal F}_{\rm He}}{2} \left(1+ \tanh \left[ \frac{y(z_{\rm{reio}})-y(z)}{\Delta_y} \right] \right),
\label{eqn:tanh}
\end{equation}
where ${\cal F}_{\rm He}=n_{\rm{HeII}}/n_{\rm{H}}$ is the ratio of singly ionized Helium to Hydrogen atoms\footnote{The contribution from doubly ionized Helium to the free electron fraction is added in the form of another hyperbolic tangent but at lower redshift, $z\sim 3.5$, when HeII is expected to be ionized \cite{Planck:2018vyg}.}, 
$y(z)=(1+z)^{\gamma}$, $\Delta_y=\gamma (1+z_{\rm reio})^{\gamma-1}\Delta_z$, where 
$\Delta_z$ is the width of the transition. The parameters $\Delta_z$ and $\gamma$ are the are fixed to 0.5 and 3/2 respectively. The only free parameter that we will vary here is the reionization redshift $z_{\rm{reio}}$. With such a reionization model, Planck 2018 temperature and polarization data gives rise to an optical depth to reionization
\begin{equation}
\tau_{\rm Pl}=0.054 \quad {\rm with}\quad \sigma_{\rm  Pl}(\tau)=0.007  
\label{eq:tauPl}
\end{equation}
 where $\sigma_{\rm  Pl}(\tau)$ denotes the 68\% CL error~\cite{Planck:2018vyg}. This implies a mid-point redshift of reionization $z_{\rm{reio}} = 7.68 \pm 0.79$ at 68\% CL, suggesting that the
Universe was fully reionized by $z \simeq 6$. The interest in this model is justified by the fact that it is easy to explore
 a large set of reionization histories by varying $z_{\rm reio}$
 or even the reionization width. In the left panel of Fig.~\ref{fig:comparison_PUCH_FG}, the continuous curves illustrate the ionized fraction evolution within an hyperbolic tangent model assuming $z_{\rm reio
}=7.68$. The blue curve assumes a negligible energy
injection from DM decays and is in agreement with Planck 2018 data. The yellow curve is obtained with a larger coupling to photons affecting  the ionization history at $z\gtrsim z_{\rm reio}$. 

 \subsubsection{Reionization from stars}
 \label{sec:astro}

Apart from the hyperbolic tangent model,
in this paper, we shall also consider two explicit reionization models from 
Puchwein et al.~\cite{Puchwein:2018arm} and 
Fauchere-Gigu\`ere~\cite{Faucher-Giguere:2019kbp}, that we denote by PUCH and
FG, respectively, for short.\footnote{The PUCH model is  implemented by default in
{\tt DarkHistory} while we have  implemented the FG model by making use of the tabulated photoheating and
photoionization rates provided \hyperlink{https://galaxies.northwestern.edu/uvb-fg20}{here}. }
Those reionization scenarios rely on observations of the UV and X-ray background emission from galaxies  to model the photoionization ($\Gamma_\text{X}^{\gamma\text{-ion}}$) and photoheating (${\cal H}_\text{X}^{\gamma\text{-heat}}$) rates from astrophysical sources contributing to the $\dot{Y}^\text{astro}$ term of Eq.~(\ref{eq:dotY}) as~\cite{Liu:2019bbm,Liu:2020wqz}
\begin{eqnarray} \label{eq:Astro contributions}
	\left(\begin{matrix} \dot{T}_m^{\text{astro}} \\ \dot{x}_\text{X}^{\text{astro}}  \end{matrix} \right) = 
	\left(\begin{matrix} \frac{2}{3\left(1+{\cal F}_\text{He}+x_e\right)n_\text{H}}  \sum_X {\cal H}_\text{X}^{\gamma\text{-heat}}  \\ \subt{x}{X}  \Gamma_\text{X}^{\gamma\text{-ion}}    \end{matrix} \right) \,
\end{eqnarray}
where  X=\{HII, HeII, HeIII\}. 

  There are several differences between the PUCH and FG models.  First, the
  onset of reionization, $z_{\rm A}^{\rm max}$, is given by
  $z^{\rm max}_{\rm PUCH}=15.1$ and $z^{\rm max}_{\rm FG}=7.8$ for the
  PUCH and FG reionization models, respectively. Moreover, in the  FG model
  reionization is relatively rapid compared to the PUCH model. 
  These differences are illustrated in Fig.~\ref{fig:comparison_PUCH_FG}, where we depict the redshift evolution of  $x_e$ (left panel) and $T_m$ (right panel) with a dashed line for the FG model and a dot-dashed line for the PUCH model.  
 Let us emphasize that, for redshifts above $z_{\rm A}^{\rm max}$ and fixed values of the coupling to photons,  all $x_e(z)$ and $T_m(z)$ curves are identical by construction. Indeed, the $\dot Y^{\rm astro}$ term only accounts for extra energy injection from stars at $z<z_A^{\rm max}$. 

In the left panel of Fig.~\ref{fig:comparison_PUCH_FG}, where 
we show the evolution of the free electron fraction as a function of the redshift, we consider a DM particle with $95$~eV mass and two distinct couplings to photons (see Eq.~(\ref{eq: Axion decay rate})), $g_{a\gamma \gamma}=10^{-15}$~GeV$^{-1}$ (blue lines) and $g_{a\gamma \gamma}=10^{-13}$~GeV$^{-1}$ (orange lines), that correspond, respectively, to a negligible and a significant DM energy injection.
After implementing the PUCH and FG models in {\tt DarkHistory} and using Eq.~(\ref{eq:tau}) to evaluate the optical depth, we obtain
for the default PUCH and FG reionization models (i.e. the blue curves with negligible DM
energy injection):~\footnote{The very same results can be
obtained by implementing these two reionization histories in {\tt
  CLASS} with the {\tt reio\_parametrization} set to {\tt reio\_inter}
that takes into account tabulated values of $(z,x_e)$ between $z=6$
and $z^{\rm max}_{\rm PUCH}=15.1$ and $z^{\rm max}_{\rm FG}=7.8$ for
PUCH and FG reionization models explicitly. Notice that our result for $\tau_{\rm FG}$ differs from the reported value by \cite{Faucher-Giguere:2019kbp} by 0.002. This might be due to a slight difference in the prescription for computing $\tau$.}
\begin{equation}
\tau_{\rm PUCH}=0.064~,\quad {\rm and} \quad
\tau_{\rm FG}=0.052~.\label{eq:tauFGP}
\end{equation}
Comparing these optical depths to the one reported by Planck in
Eq.~(\ref{eq:tauPl}), it is clear that the FG reionization model
will lead to more conservative bounds on the DM scenario than the PUCH model. Indeed,  the latter gives rise to a
larger optical depth to reionization leaving less room for an extra DM contribution to the free electron fraction. 

The IGM temperature evolution is depicted in the right panel of
Fig.~\ref{fig:comparison_PUCH_FG}. 
  We show in blue (orange) an ALP-photon coupling of 
$g_{a\gamma \gamma}=10^{-15}$~GeV$^{-1}$ ($g_{a\gamma \gamma}=10^{-13}$~GeV$^{-1}$) and the dashed and dot-dashed curves show the IGM temperature evolution in the PUCH and FG models. We  clearly see the differences between these two reionization models as well as the impact of  DM energy injection.    In all cases, the DM decay into photons  induces higher IGM temperature for larger couplings to photons (well visible for $z>z^{\rm max}_{\rm PUCH})$.
Also, for both PUCH and FG models, we see the presence of a bump in the IGM temperature that roughly starts at the onset of reionization and peaks when the latter is completed. 
In the PUCH model, the changes in the IGM temperature are smoother as reionization starts at higher redshifts than in the FG case. In the FG model, reionization happens on a much shorter time scale, the changes are more abrupt and cause a sharper peak in the evolution of the matter temperature at reionization ($z\simeq 8$). Notice though that the values of $T_m$ on this peak of temperature remain at most within a factor $\sim 2$  from the values of $T_m$ in the $z<6$ redshift range, where current observations of the IGM temperature are relevant.

\section{CMB analysis}
\label{sec: CMB analysis}

As discussed in the previous sections, energy injection from annihilations or decays of
DM particles in the early universe can leave a detectable imprint on CMB anisotropies, see
e.g.~\cite{Shull:1985,Chen:2003gz,Slatyer:2009yq,Slatyer:2015kla}, that can lead to strong constraints on beyond the Standard Model scenarios, see
e.g.~\cite{Padmanabhan:2005es,Cadamuro:2011fd,Lopez-Honorez:2013cua,Diamanti:2013bia,Liu:2020wqz,Slatyer:2016qyl,Bolliet:2020ofj}. 
The energy injection efficiencies and the different reionization histories, discussed in Secs~\ref{sec:DM} and~\ref{sec:reio}, are  two  crucial inputs for our CMB analysis. In this section, we first discuss in  Sec.~\ref{sec:CLASS} how in practice we deal with energy injection from  DM and stars in the public CMB Boltzmann solver code {\tt CLASS}. Based on this approach, we then present in Sec.~\ref{sec:CMBcons} the results of a Monte Carlo Markov Chain (MCMC) analysis that provides updated constraints on the DM parameter space using the CMB temperature and polarization measurements by Planck 2018. Finally, we end the section with the prospects to constrain the DM coupling to photons with future CMB measurements by using the forecasted sensitivities in the measurement of the optical depth to reionization.

\subsection{Energy deposition from ${\rm DM} \to \gamma\gamma$ in {\tt CLASS}}
\label{sec:CLASS}
The {\tt CLASS} Boltzmann solver~\cite{CLASS,Blas:2011rf}  can account for exotic energy injection at high redshifts ($z\gtrsim z_{\rm reio}$) building upon the {\tt ExoCLASS} extension, see Refs.~\cite{Stocker:2018avm,Lucca:2019rxf}. In the case of DM decays, the default implementation in the {\tt injection} module fixes the energy deposition efficiencies $f_c(x_e, z)$ to those given in Ref.~\cite{Chen:2003gz}, which essentially reduce to $f_c(x_e, z)=1/3$ for $c=$HII, heat and exc at large $z$. At $z\lesssim z_{\rm reio}$ the ionized fraction follows by default the hyperbolic tangent model presented in Sec.~\ref{sec:tanh}.  The {\tt thermodynamics} module  allows however to implement any reionization history by providing a list of $x_e(z)$ points between which {\tt CLASS} interpolates.

In order to efficiently account for a more accurate  treatment  of energy deposition from dark matter and stars, we have made slight modifications of both the {\tt injection} and {\tt thermodynamics} modules of {\tt CLASS}:\footnote{Our modified version of {\tt CLASS} is available  at the following link: \url{https://github.com/llopezho/CLASS_DMdecay}.}
\begin{itemize}
    \item At $z\lesssim z_{\rm A}^{\rm
  max}$, we account for specific reionization from stars  (PUCH or FG models) interpolating, within the {\tt thermodynamics} module,  a tabulated evolution of $x_e(z)$  between $z=6$ and $z_{\rm A}^{\rm
  max}$ for different values of the DM parameters $m_a$ and $g_{a\gamma\gamma}$  within the ranges of interest, see Sec.~\ref{sec:CMBcons}. These tabulated values have been obtained with {\tt DarkHistory} and take into account the convoluted effect of  DM decay and reionization from stars. 
\item Before reionization,    we have made use of an approximation to the energy deposition, described in Eq.~(\ref{eq:dep}), that facilitates the computation of energy injection efficiencies for any DM mass and couplings relevant here. We discuss the latter in more detail below. Let us also mention that in all cases we have made use of the default {\tt HyRec} recombination algorithm \cite{Ali-Haimoud:2010tlj,Lee:2020obi}
\end{itemize}

\begin{figure}[t]
    \centering
\includegraphics[width=0.45\linewidth]{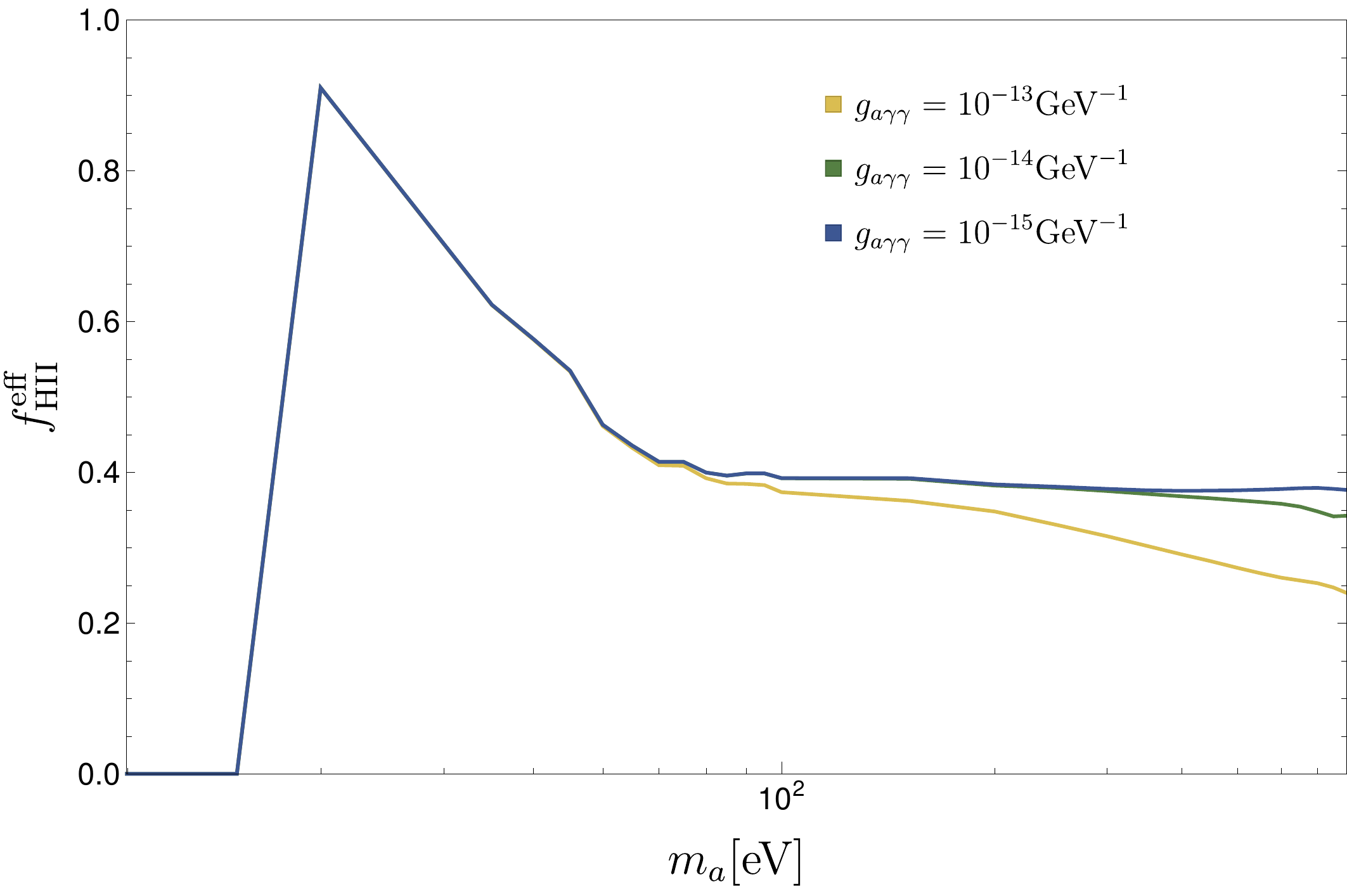} \includegraphics[width=0.45\linewidth]{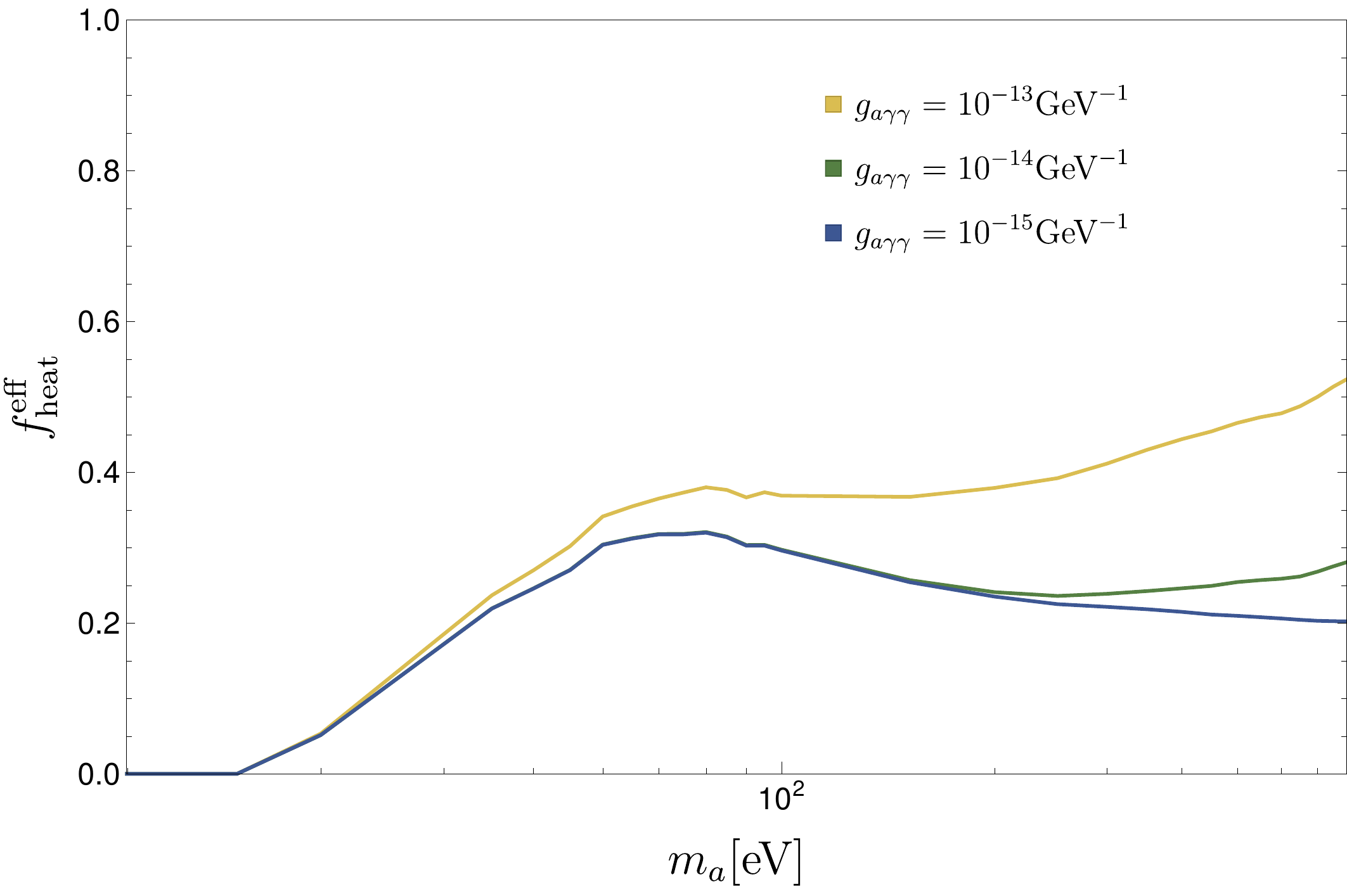}

\caption{ \textbf{\textit{ Effective energy deposition efficiencies}} dependence on the DM  mass  and coupling to photons. The plots illustrate the case of energy deposition  into Hydrogen ionization ($c=$ HII, left panel)  and  heating of the IGM ($c=$ heat, right panel) as a function of $m_a$, in the mass range considered in our analysis, for  couplings  $g_{a\gamma\gamma}= 10^{-15}$ (in blue), $10^{-14} $ (in green) and $10^{-13}$~GeV$^{-1}$ (in orange). Those  efficiencies are  used in our CMB analysis at $z>z_{A}^{\rm max}$, see text for details.}.
    \label{fig:fceff}
\end{figure}

  It is well known that in the case of dark matter decays, efficient energy
deposition is delayed to later times  with respect to e.g. the annihilating DM case, see for example  the discussion
in Refs.~\cite{Padmanabhan:2005es,Slatyer:2015kla,Diamanti:2013bia,Slatyer:2016qyl,Liu:2020wqz}. In Ref.~\cite{Slatyer:2016qyl}, it was shown  by means of a
principal component analysis that the impact of DM decays on the CMB (between reionization and recombination) is well captured using the energy deposition efficiencies $f_c(x_e,z)$ at redshift $z \simeq 300$, as expected from the results of Ref.~\cite{Finkbeiner:2011dx}. This allows to shortcut the treatment of high redshift energy deposition by using:
\begin{equation}
    \left(\frac{dE_c (x_e,z)}{dt\,dV}\right)_{\rm deposited} =   f_c^{\rm eff} \, \left(\frac{dE (x_e,z)}{dt\,dV}\right)_{\rm injected}\, \quad {\rm for} \quad z>z_{\rm A}^{\rm max}\, ,
    \label{eq:depeff}
\end{equation}
where $f_c^{\rm eff}=f_c(x_e,z=300)$ is used as an effective energy deposition efficiency parameter. In our CMB analysis, we use this  approximation at high redshifts instead of the full $f_c(x_e,z)$ treatment of  Eq.~(\ref{eq:dep}). 
In Fig.~\ref{fig:fceff}, we illustrate the dependence of $f_{\rm HII,heat}^{\rm eff}$ on the dark matter mass for different values of the  coupling to photons (or equivalently of lifetimes). We see that for $m_a\lesssim {\cal O } (100)$  eV, ionization becomes the main channel for energy deposition at large redshifts, except below $m_a<26$ eV where the excitation channel is dominant.

\begin{figure}[t]
    \centering
\includegraphics[width=0.45\linewidth]{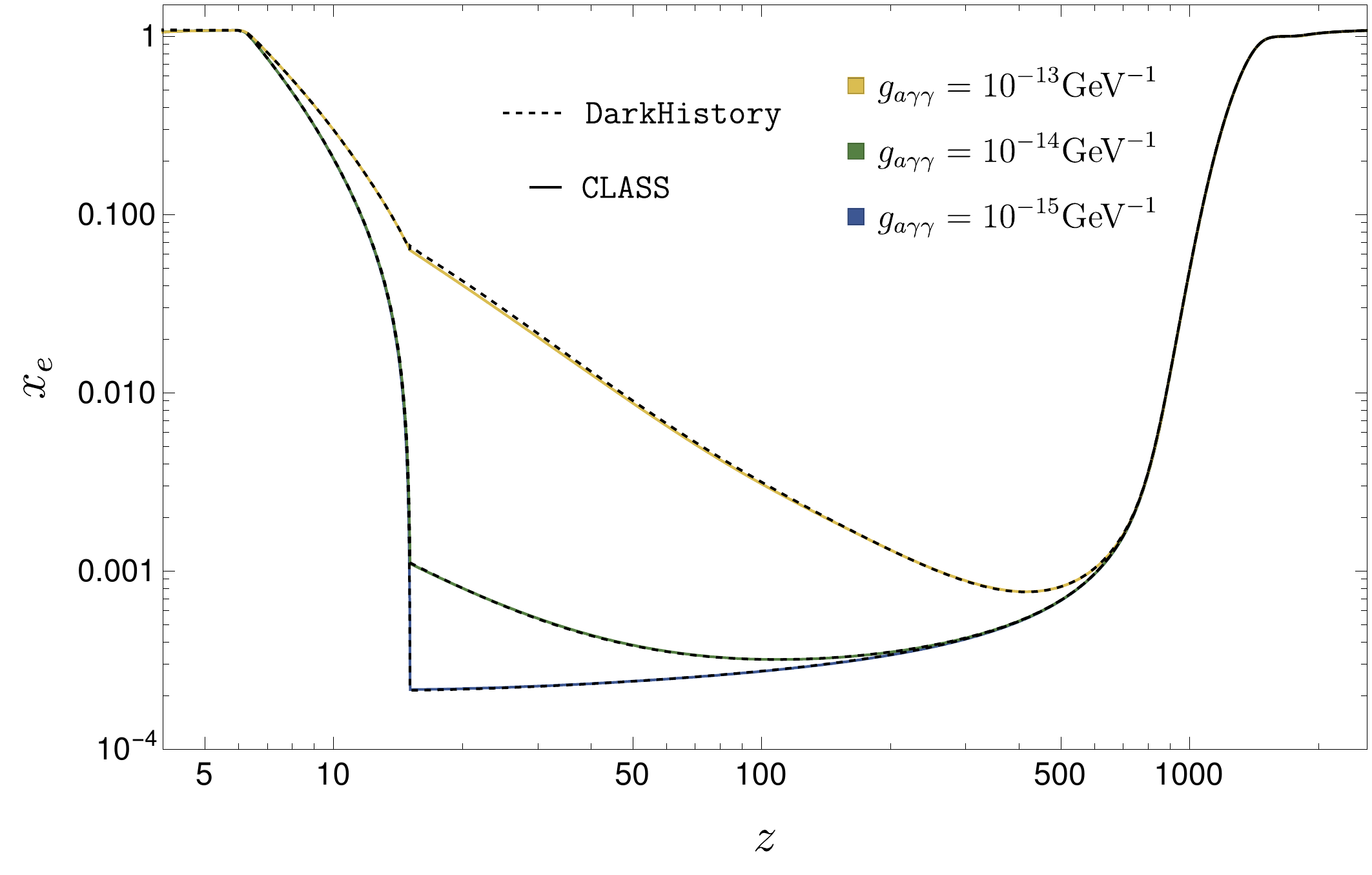}

\caption{\textit{\textbf{Comparison of the different numerical approaches}} when considering the dark matter energy injections through decays of a dark matter particle of mass $m_a=~95$ eV and different couplings and reionization models. On the left panel, we depict the ionization history when considering a PUCH reionization model in our modified version of the {\tt CLASS} Boltzmann package code (dashed lines) and that recovered from the full treatment of the {\tt DarkHistory} package (continuous colored lines) for couplings  $g_{a\gamma\gamma}= 10^{-15}$ (in blue), $10^{-14} $ (in green) and $10^{-13}$~GeV$^{-1}$ (in orange).}
    \label{fig:ion}
\end{figure}

In Ref.~\cite{Slatyer:2016qyl}, it  was explicitly checked that $f_c^{\rm eff}$ is in excellent agreement with the first principal component of $f_c(x_e,z)$    for decaying DM masses above $10^4$ eV. 
Making use of the {\tt DarkHistory} package and of our modified {\tt CLASS} code, we found an excellent agreement on $x_e(z)$ when comparing the effective or the  full energy deposition approaches for $z_{\rm reio}<z<10^3$ and  dark matter masses between 20.4 eV and $10^4$ eV. This is illustrated in Fig.~\ref{fig:ion} where we focus on a dark matter particle with a mass $m_a=95$ eV decaying into two photons with a coupling $g_{a\gamma\gamma}$ between $10^{-13}$~GeV$^{-1}$ (orange lines) and $10^{-15}$~GeV$^{-1}$ (blue lines) and assuming a PUCH reionization model. The continuous colored lines are obtained with the {\tt DarkHistory}  software using the full treatment of  $f_c(x_e,z)$, as in  Eq.~(\ref{eq:dep}), while the dashed  lines are obtained with the {\tt CLASS} code, making use of the effective energy deposition of Eq.~(\ref{eq:depeff})  with $f_c^{\rm eff}= f_c(x_e,z=300)$ from {\tt DarkHistory}. Notice that continuous and dashed lines are almost identical as expected.

\subsection{Current and future constraints for different reionization models}
\label{sec:CMBcons}

Based on the prescription for energy injection at recombination and reionization described in Sec.~\ref{sec:CLASS}, we now use Planck 2018 data to derive constraints on sub-keV decaying dark matter. 
We present the bounds in the plane of the DM mass $m_a$  and DM coupling to photons $ g_{a\gamma\gamma}$, that effectively set the decay rate (see Eq.~(\ref{eq: Axion decay rate})). We focus on the mass and  coupling ranges:
\begin{equation}
   m_a \supset [10, 10^4] \, {\rm eV}\quad {\rm and} \quad \log_{10}[g_{a\gamma\gamma}\times {\rm GeV}] \supset [-12, -16]~.
   \label{eq:ranges}
\end{equation}
We also analyse the impact of the underlying  reionization model on the constraints. More precisely, we derive the bounds that arise in  the case of the hyperbolic tangent reionization model of Sec.~\ref{sec:tanh}, denoted by $\tanh$ for short, and compare them to the explicit FG and PUCH astrophysical models presented in Sec.~\ref{sec:astro}.

\begin{figure}[t]
    \centering
\includegraphics[width=0.75\linewidth]{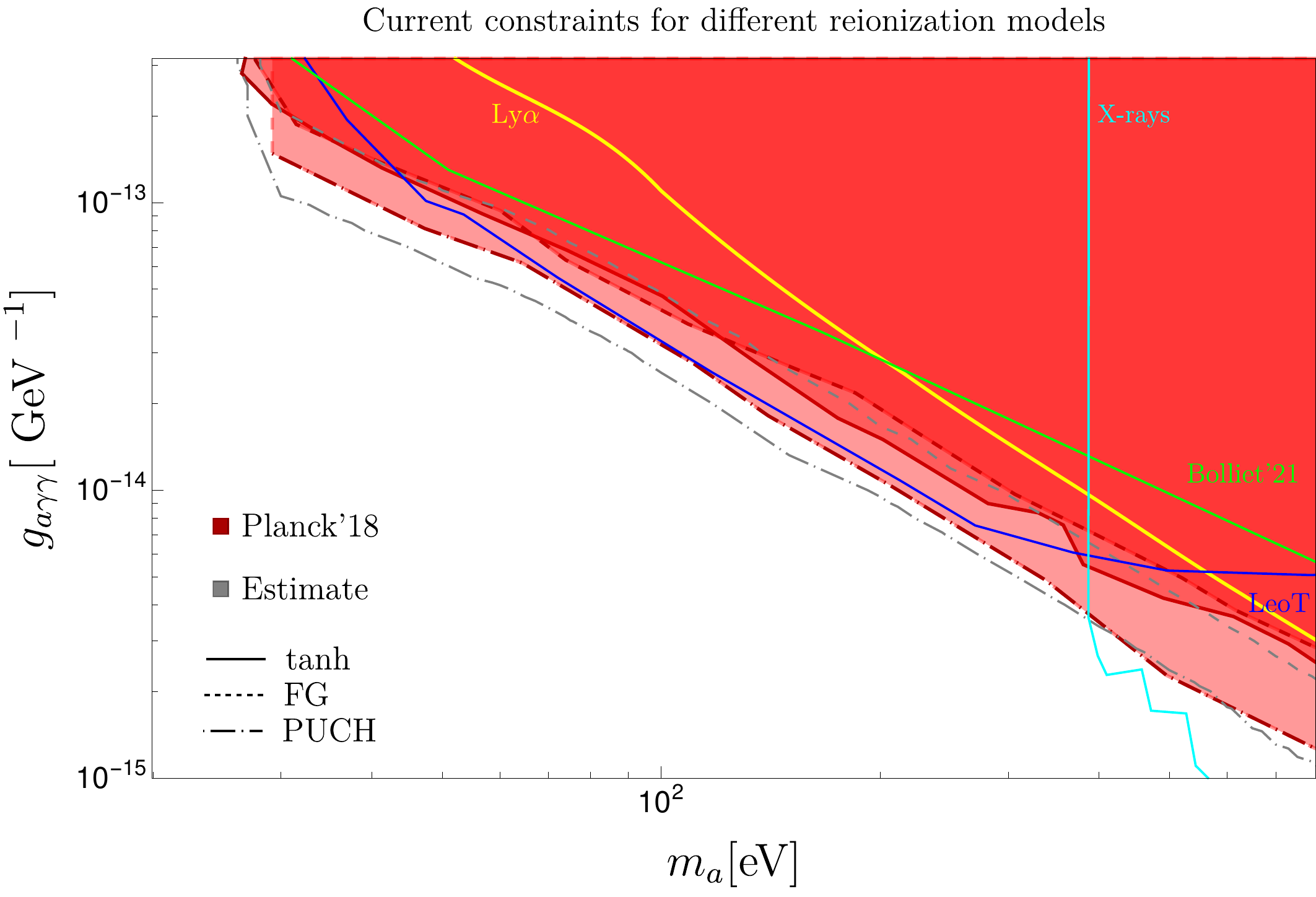}
\caption{\textit{\textbf{Exclusion limits from CMB anisotropies}} in the $(m_{a} [{\rm eV}],g_{a\gamma\gamma}[1/{\rm GeV}])$ plane. The red lines correspond to the regions excluded at 99\%~CL from Planck 2018 data for  different reionization histories: the standard hyperbolic tangent description (continuous), the Fauchere-Gigu\`ere (FG) model (dashed) and the Puchwein (PUCH) model (dot-dashed). In the case of FG and PUCH models, the corresponding gray lines 
show a rough estimate of the exclusion limits based on the evaluation of the optical depth to reionization. The yellow continuous line represents the most stringent constraint derived in Sec.~\ref{sec:Lycons} from Lyman-$\alpha$ data assuming a $\tanh$ reionization. The other continuous colored lines correspond to existing limits from  a previous CMB analysis~\cite{Bolliet:2020ofj}  (green), X-ray analysis~\cite{Cadamuro:2011fd} (cyan) as well as the conservative constraint from Leo-T~\cite{Wadekar:2021qae} (blue).}
    \label{fig:reio-cons}
\end{figure}

Before going through the detailed statistical analysis, we can infer a first rough estimate of the expected bounds 
in the case of the PUCH and FG models.
To that purpose, we compute with our modified version of {\tt CLASS} the optical depth to reionization  over the whole range of $(m_a,g_{a\gamma\gamma})$ reported in \ref{eq:ranges} and  estimate the bounds by excluding the region where $\tau> \tau_{\rm Pl}+2\times \sigma_{\rm Pl}$.  The corresponding limits are shown with dashed and dot-dashed gray lines in
Fig.~\ref{fig:reio-cons} for the  FG and PUCH reionization models, respectively. The expected excluded regions at $2\sigma$~CL lie above those
gray lines. 
Notice that  the PUCH model gives rise to stronger constraints, as anticipated in
Sec.~\ref{sec:astro}. Indeed even without dark matter energy injection, the PUCH model yields an optical depth, $\tau=\tau_{\rm PUCH}$, that is more than $1\sigma$ above the central value preferred by the Planck 2018 data. Including DM energy injection, the limit on  $g_{a\gamma\gamma}$ for fixed $m_a$ in the PUCH model is roughly half an order of magnitude stronger than in the FG case.

We can now perform a full Monte Carlo analysis. The minimal set of cosmological parameters considered in our analysis includes:
\begin{equation}
    \{\Omega_b h^2,\Omega_a h^2, 100\theta_*, \ln[10^{10}A_s], n_s,\log_{10}[m_a/ {\rm eV}], \log_{10}[g_{a\gamma\gamma}\times {\rm GeV}]\}\,.
    \label{eq:cosmoset}
\end{equation}
In the case of the PUCH and FG reionization models we work with fixed photoionization and photoheating rates and thus perform the MCMC on the set of  parameters ~(\ref{eq:cosmoset}). In contrast, in the case of the  hyperbolic tangent  model, the set of parameters is supplemented by the reionization redshift $z_{\rm reio}$. In the latter case, we can thus effectively marginalize over multiple reionization scenarios. 
In Eq.~(\ref{eq:cosmoset}), $\Omega_b h^2$
and $\Omega_a h^2$ are the relative baryon and decaying dark matter
densities today, $\theta_*$ is the acoustic scale angle and $A_s$ and $n_s$ are, respectively, the 
amplitude and spectral index of the primordial power spectrum. For the latter purposes, we have run the {\tt MontePython} software~\cite{Brinckmann:2018cvx}
interfaced with our modified version of {\tt CLASS} and used the baseline
TT, TE, EE + lowE Planck 2018 likelihoods.  The resulting bounds at 99\% CL are
depicted in Fig.~\ref{fig:reio-cons} in thick red continuous, dashed  and dot-dashed lines for, respectively, the $\tanh$, FG and PUCH reionization models. 
Interestingly, we notice that the results of the Monte Carlo analysis are
in good agreement with the estimated bounds (gray
lines). In addition, we note that the tanh model, marginalizing over the reionization redshift in the range $z_{\rm reio}= 5$ to 13, leads to a constraint on the parameter space that is very similar to the conservative
case of a FG reionization scenario. We also note that we have not found any clear degeneracy between the DM parameters and any of the cosmological parameters in the analysis.

The CMB bounds derived here are more stringent than the previous ones from~\cite{Cadamuro:2011fd,Bolliet:2020ofj}. Indeed, our  analysis  differs from the previous ones  in a few aspects. First, we use  a more recent CMB data release,  which translates into a lower value of $\tau$. Second, we make use of more accurate values for the energy deposition efficiency coefficients by including the $f_c(x_e,z=300)$ computed from {\tt DarkHistory} and we exploit the full CMB anisotropy spectrum information rather than just the optical depth to reionization. Also, we perform a full MCMC analysis to extract the constraints from CMB anisotropies. Let us also emphasize that our bounds are competitive  with the constraints from the radiative cooling gas rate of the Leo-T dwarf galaxy~\cite{Wadekar:2021qae} in the case of the more aggressive PUCH reionization scenario.

\begin{figure}[t]
    \centering
\includegraphics[width=0.75\linewidth]{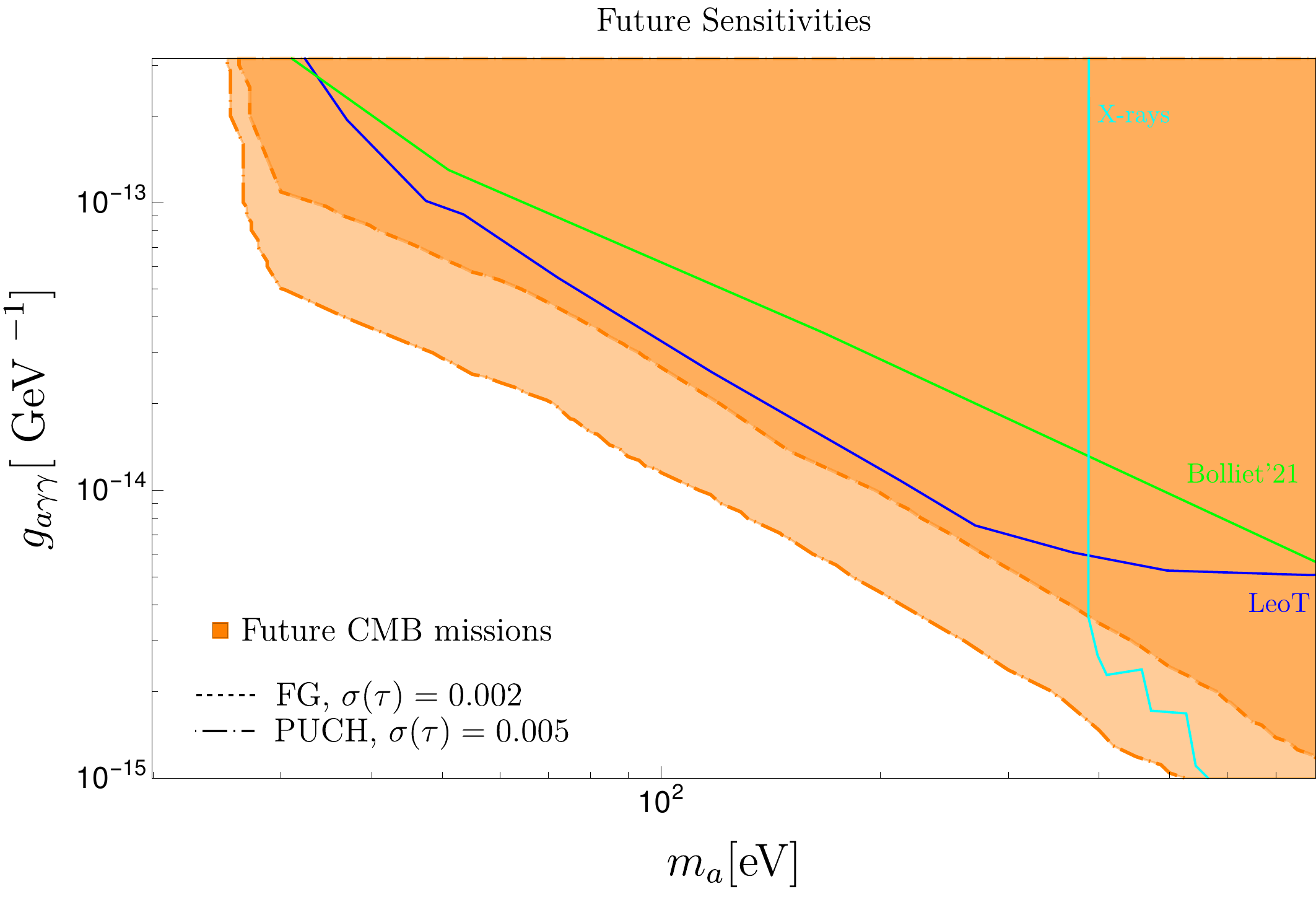}
    \caption{\textit{\textbf{Sensitivities of future CMB missions. }} The orange lines correspond to future bounds that could be reached in the $(m_{a} [{\rm eV}],g_{a\gamma\gamma}[1/{\rm GeV}])$ plane assuming a fiducial PUCH (dot-dashed) (FG (dashed))  reionization model and a 1$\sigma$ error on $\tau$ reduced to 0.005 (0.002). Other continuous colored lines  correspond to existing limits from a conservative Leo-T analysis~\cite{Wadekar:2021qae} (blue), a previous CMB analysis~\cite{Bolliet:2020ofj}  (green) and X-ray limits \cite{Cadamuro:2011fd} (cyan). }
    \label{fig:reio-cons-fut}
\end{figure}

Concerning future prospects, CMB-S4 surveys are expected to reach a $1\sigma$ uncertainty on the  optical depth to reionization of $\sigma(\tau)=0.0025$~\cite{Abazajian:2019eic}. Preliminary estimates also
show that by combining measurements of the kinematic Sunyaev-Zeldovich (kSZ) effects with the CMB-S4 data, the sensitivity could be improved and reach $\sigma(\tau)=0.002$, very close to the cosmic
variance limit (CVL)~\cite{Abazajian:2019eic}. These values have to be compared to 
$\sigma_{\rm Pl}(\tau)=0.007$ from Planck 2018~\cite{Planck:2018vyg}. One can then estimate how the constraints shown
in Fig.~\ref{fig:reio-cons} would improve with future CMB experiments by considering the improved sensitivities on the determination of the optical depth.  Here we impose $\tau<\tau_{\rm Pl}+ 2\times \sigma_{\rm fut}(\tau)$, i.e. assuming that the central value of $\tau$ would not change but the error would be decreased to $\sigma_{\rm fut}(\tau)<\sigma_{\rm Pl}(\tau)$. The resulting forecasts are shown in Fig.~\ref{fig:reio-cons-fut}.  Considering
$\sigma_{\rm fut}(\tau)=0.002$,  the CMB bound assuming a FG reionization (dashed orange line) could become at least as strong as the
current PUCH limit with $\sigma_{\rm Pl}(\tau)=0.007$.
This implies that with CMB-S4 \& kSZ,
the bound arising from CMB anisotropies could become as good as the one from the Leo-T gas temperature~\cite{Wadekar:2021qae}, even in the more conservative reionization model considered here (FG). We have also checked that the CVL relative uncertainty $\sigma(\tau)/\tau=2.5\%$~\cite{Watts:2018etg} does not lead to significant change in the limit and the resulting sensitivity is essentially superposed to the $\sigma(\tau)=0.002$ case.  Very interestingly, in the case of
a reionization model such as the PUCH one, basically any improvement in
the precision of the optical depth to reionization will 
improve upon the Leo-T bound. Furthermore, we also show in Fig.~\ref{fig:reio-cons-fut} the estimate of the limit  for a modest improvement from
$\sigma_{\rm Pl}(\tau)=0.007$ to $\sigma_{\rm fut}(\tau)=0.005$ with a
dot-dashed orange line. Such a small improvement would increase the CMB bound on 
$g_{a\gamma\gamma}$  by almost one order of magnitude and become the most stringent bound on this mass range.

\section{Lyman-$\alpha$ constraints}
\label{sec:Lycons}

In the previous section, we have used CMB anisotropies to constrain the effects of the DM energy injections  on the ionization history. However, and as aforementioned, DM decays also affect the IGM temperature, see e.g Fig.~\ref{fig:comparison_PUCH_FG} (right panel). Using recent determinations of the IGM temperature in the redshift range $3.6<z<5.8$ from Lyman-$\alpha$ data~\cite{Walther:2018pnn,Gaikwad:2020art}, the authors of Ref.~\cite{Liu:2020wqz} derived constraints on the mass and coupling to photons of DM particles with masses above $10$ keV. They have  used the {\tt TIGM} branch of the {\tt DarkHistory} code, where they implemented a modified chi-square test that only penalizes temperature histories that overheat the IGM compared to the data. 
In this section, we extend such an analysis down to DM masses of 30 eV,  using our modified version of the {\tt TIGM} branch of the {\tt DarkHistory} code.~\footnote{We have modified  all branches of the {\tt DarkHistory}  code to include  collisional excitation processes at higher redshifts, before the onset of reionization, see  the discussion in appendix~\ref{sec:modif_darkhistory}.}

The analysis of Ref.~\cite{Liu:2020wqz} includes  conservative assumptions concerning the astrophysical sources of heating and ionization. On the one hand,  the astrophysical source for photoheating   is set to zero (${\cal H}_\text{X}^{\gamma\text{-heat}}=0$ in Eq.~(\ref{eq:Astro contributions})) whereas a  minimal astrophysical HI photoionization rate,  denoted by $\Gamma_\text{HI}^{\gamma\text{-ion}}$  in Eq.~(\ref{eq:Astro contributions}), is considered.~\footnote{This branch of the code also neglects the ionization of HeII to HeIII. This is justified for redshifts prior to the full ionization of HeII ($z \sim 3$ \cite{Becker:2010cu}) which is the case in this section where all data points are at redshifts above $z\sim 3.6$.} 
The latter is obtained by requiring that all the contributions to the ionized fraction sum up, at small redshifts (from the onset of reionization until today) to the hyperbolic tangent model discussed in Sec.~\ref{sec:tanh} with a $z_{\rm reio}$ within $1\sigma$ of the central value of Planck 2018 data.~\footnote{In \cite{Liu:2020wqz} two different parametrizations of the ionization fraction were considered: the $\tanh$ model and to the so-called FlexKnot parameterization that is also used in the Planck analysis \cite{Planck:2018vyg}. We verified that our bounds are similar for both models so, for simplicity, we restrict ourselves to the $\tanh$ case in this work.}
 In practice, one imposes 
\begin{eqnarray}	
\dot{x}_\text{HII}^\text{astro} =  
\frac{\dot{x}_{\text{e}}^{\tanh}}{1+\mathcal{F}_{\text{He}}}- 
\subt{\dot{x}}{HII}^\text{\rm DM} - 
\subt{\dot{x}}{HII}^\text{(0)} \, ,
\end{eqnarray}
where it has been assumed that Hydrogen and Helium have similar ionized fractions so that $x_\text{HII}^{\tanh}= x_e^{\tanh}/(1+{\cal F}_\text{He})$, with ${\cal F}_{\rm He}=n_{\rm{HeII}}/n_{\rm{H}}$ the ratio of singly ionized Helium to Hydrogen atoms.

\begin{figure}[t]
	\centering
	\includegraphics[width=0.48\linewidth]{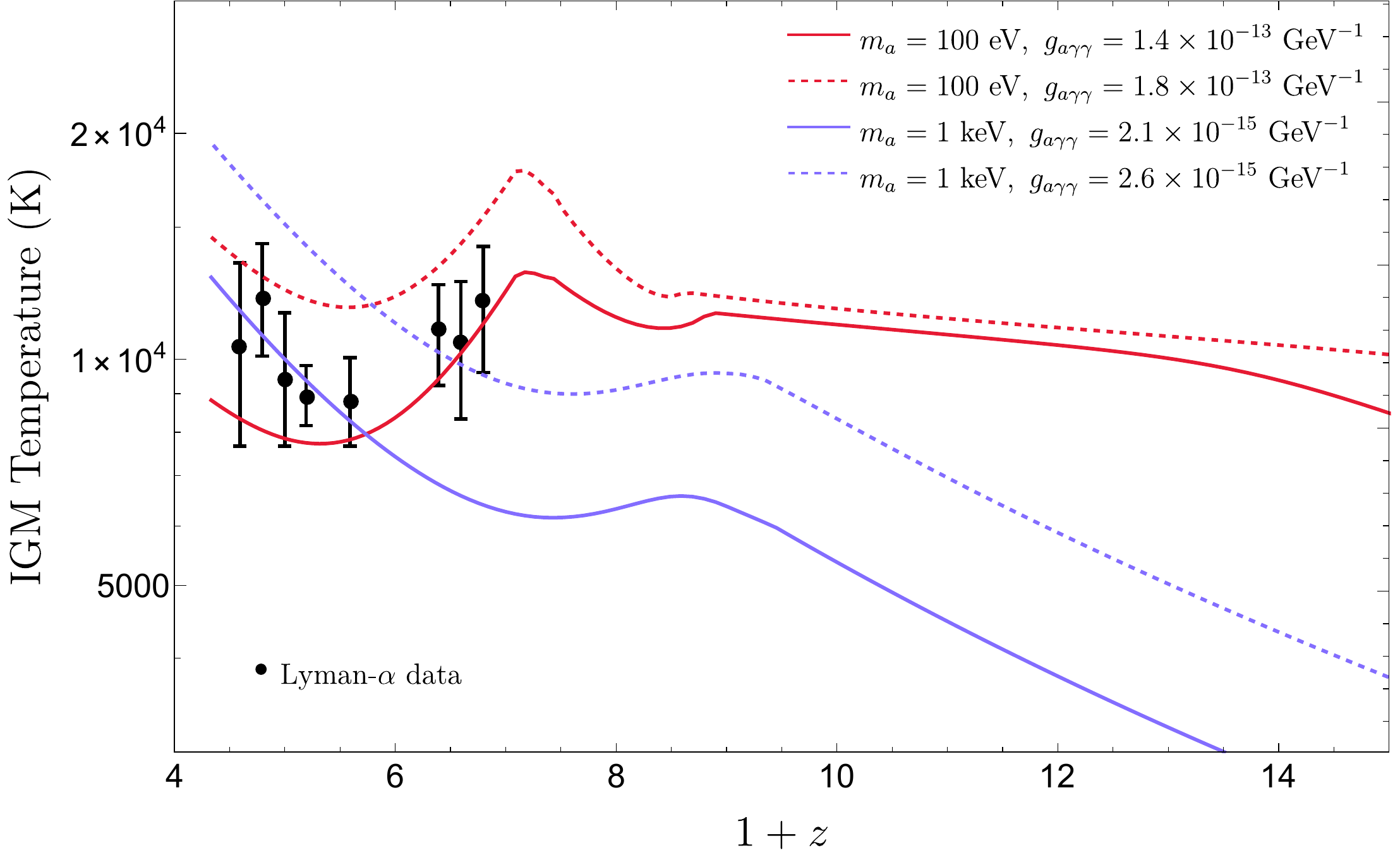} \, \,
	\includegraphics[width=0.48\linewidth]{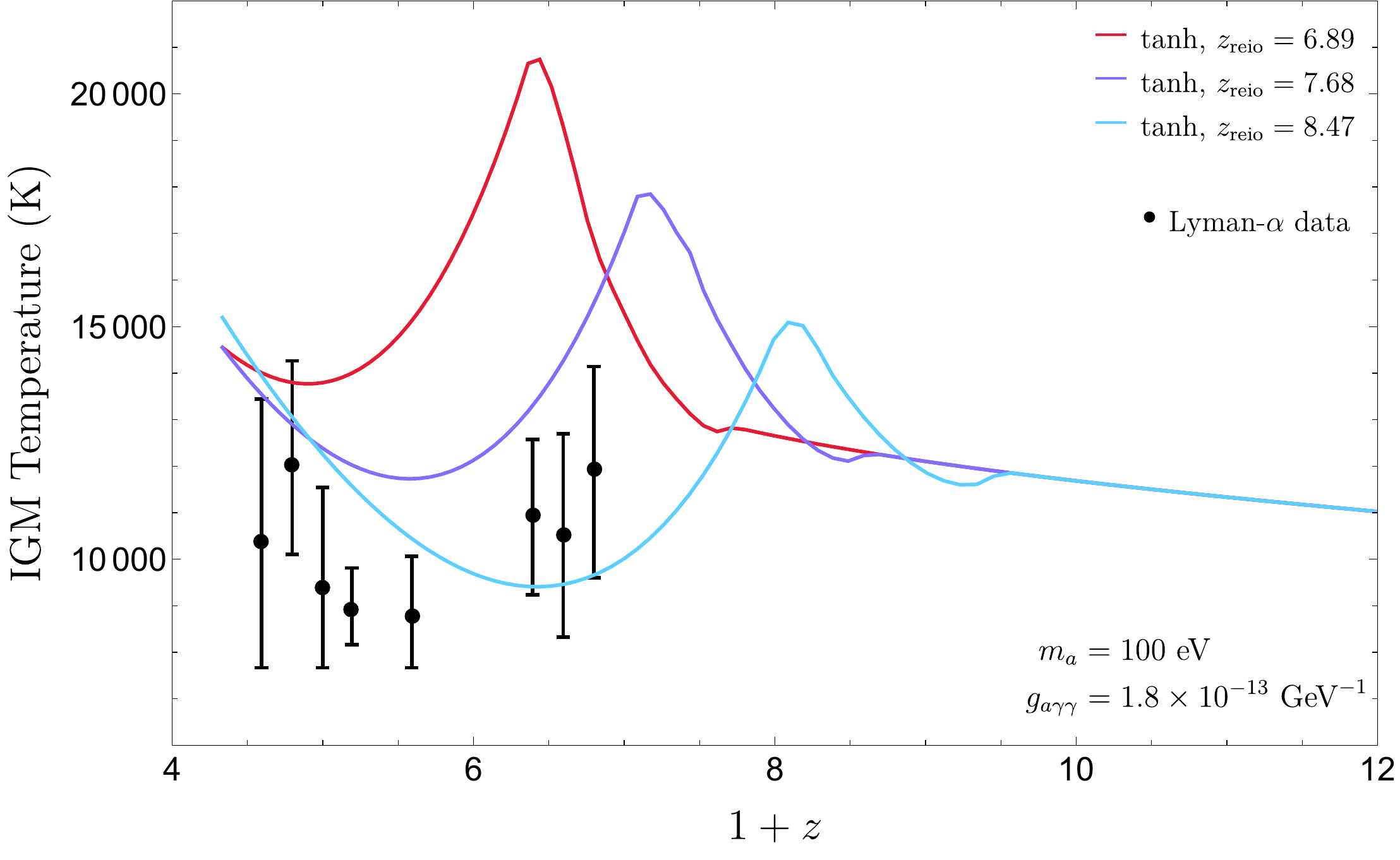}
	\caption{\textit{\textbf{Evolution of the IGM temperature.}} Left panel: Solid lines depict the evolution for values of the coupling to photons compatible with Lyman-alpha data from \cite{Walther:2018pnn,Gaikwad:2020art} and whereas dashed lines correspond to values incompatible with the same data. The ionization fraction has been fixed to the tanh model with $z_{\rm reio}=6.89$ Right panel: Same temperature evolution as in the left panel but for fixed DM mass and coupling to photons and three different values of $z_{\rm reio}$.}
	\label{fig:TIGM vs z}
\end{figure}

 In the left panel of Fig.~\ref{fig:TIGM vs z}, we show  the evolution of the IGM temperature for DM masses of 100 eV (red) and 1 keV (blue) in the redshift range  $3<z<14$ assuming a $\tanh$ reionization model with $z_{\rm reio}=7.68$. 
As expected, for shorter lifetimes (dashed curves) the effects of DM energy injection cause earlier and stronger heating of the IGM when compared to the case of longer lifetimes (solid curves). Following the modified chi-squared test described in the first paragraph of this section, the former cases are excluded by the IGM temperature data at 95\%CL whereas the latter ones are still compatible with the data.~\footnote{We follow the same prescription as in \cite{Liu:2020wqz} and discard two of the Lyman-$\alpha$ data points of Ref. \cite{Walther:2018pnn}.} On the other hand, when increasing the DM mass, we find that the heating starts at a later time but grows at a faster pace. 
 
   In the right plot of Fig. \ref{fig:TIGM vs z}, we fix the DM mass and lifetime and show instead the dependence of the IGM temperature evolution on the parameter $z_{\rm reio}$ that controls the time of reionization in the tanh model. We show the evolution for $z_{\rm reio}$ between 6.89 (red) and $8.47$ (cyan), corresponding to the values at $1\sigma$  around the Planck 2018 central value $z_{\rm reio}=7.68$ (blue)~\cite{Planck:2018vyg}. Note also the presence of a peak at a redshift around $z=z_{\rm reio}$. The peak occurs at a smaller redshift when reionization happens later (smaller $z_\text{reio}$). From this plot, it appears that an earlier reionization is easier to comply with the Lyman-$\alpha$ data.

\begin{figure}[t]
	\centering
	\includegraphics[width=\linewidth]{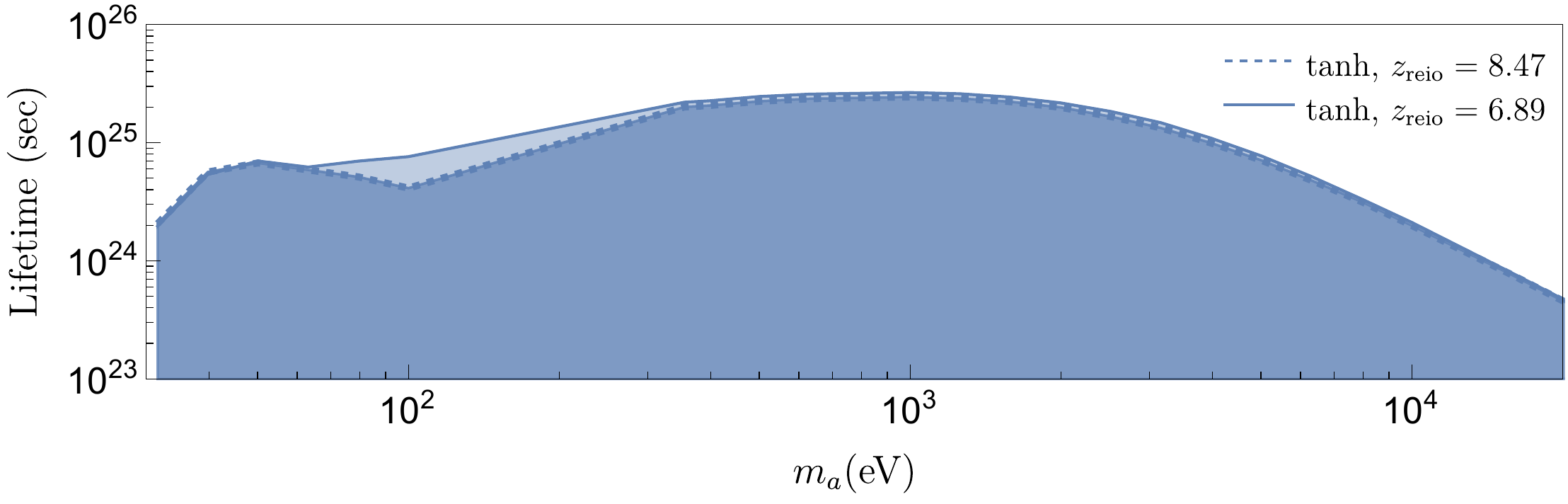}
	\caption{\textit{\textbf{Constraints on the lifetime of DM decay to two photons from IGM temperature data}} at 95\%CL assuming a tanh model for the ionized fraction. The lighter (darker) contours refers to reionization histories that start at a redshift $1\sigma$ above (below) the central value measured by Planck 2018 data.}
	\label{fig:Lifetime bound}
\end{figure}

We have then performed a systematic analysis to find the DM parameters ($m_a, g_{a\gamma \gamma}$) that are excluded by the Lyman-$\alpha$ data at 95\%CL due to overheating of the IGM temperature. 
In Fig.~\ref{fig:Lifetime bound} we present these bounds in terms of the DM mass and lifetime, which is equal to $\Gamma_{\rm dec}^{-1}$ with $\Gamma_{\rm dec}$ given by Eq.~(\ref{eq: Axion decay rate}), and their dependence on $z_{\rm reio}$. In the mass range between 30 eV to 1 keV we obtain bounds on the lifetime between $ 2\times 10^{24}$ and $ 2\times 10^{25}$ seconds. The bounds are stronger for late reionizations  and can differ by up to a factor of three, for masses $\sim 100$ eV, between the largest and smallest $z_{\rm reio}$ consider in this work. This can be traced back to the small bump   in temperature arising at the onset of reionization and that is well visible in  Fig.~\ref{fig:TIGM vs z}. There  we see that for late reionization scenarios this small increase in temperature is probed by  the Lyman-$\alpha$ data, while for earlier reionizations the bump happens at larger redshifts where there is no data yet. Furthermore, we find that the bump is more prominent for DM masses around 100 eV thus causing the biggest difference at those masses.    

Finally, in order to compare the Lyman-$\alpha$ bounds obtained here to the ones from the CMB analysis derived in the previous section,  we project in Fig.~\ref{fig:reio-cons}, the most stringent  constraint on the lifetime shown in  Fig.~\ref{fig:Lifetime bound} (for $z_{\rm reio}=6.89$) with a continuous yellow line. As we see, the very conservative assumption made here on the heating of the IGM in a $\tanh$ model  gives rise to a bound that can readily compete with the previous CMB bounds from Ref.~\cite{Bolliet:2020ofj} for $m_a\gtrsim 200$ eV. They can however not compete with the most recent  CMB bounds derived in Sec.~\ref{sec: CMB analysis}.

\section{Conclusion}
\label{sec:concl}
The ionization history and the evolution of the IGM temperature are known to be very sensitive to DM injections of energy and can be tested via CMB anisotropies and Lyman-$\alpha$ data. In this work, we have used Planck 2018 data and recent determinations of the IGM temperature at low redshifts, to update and extend previous constraints on the decay to photons of DM particles with masses between $20.4$ eV (two times the Lyman-$\alpha$ threshold) and $\sim 1$ keV (where X-ray bounds kick in, see \cite{Cadamuro:2011fd}).
To derive these constraints, we relied on the {\tt DarkHistory} v1.1 code~\cite{Liu:2019bbm} that calculates the energy deposition efficiencies as a function of the redshift, for any type of exotic injection source and deposition channel, and self-consistently computes the corresponding evolution of the Hydrogen and Helium ionized fractions and the IGM temperature. These quantities can then be used to compute constraints from CMB, Lyman-$\alpha$ or e.g. future 21-cm data.

The ionized fractions and the IGM temperature are sensitive to the astrophysical model which drives   reionization. In Sec.~\ref{sec:sec2}, we have briefly described how in practice energy injection from DM and astrophysics is implemented in {\tt DarkHistory}. We have commented on some modifications that we have introduced for our analysis. First, we take into account  cooling due to collisional excitation processes at redshifts above reionization
 (see Sec.~\ref{sec:reio} and App.~\ref{sec:modif_darkhistory}). The latter appears to have an important effect on the evolution of the IGM temperature for the DM masses and lifetimes range considered in this work. 
 Second, in order to study the dependence of the evolution of the IGM temperature and the ionized fractions on the astrophysical model for the photoheating and photoionization rates, we have considered the FG model~\cite{Faucher-Giguere:2019kbp} in addition to the default $\tanh$ and PUCH~\cite{Puchwein:2018arm} models implemented in {\tt DarkHistory}. 
 While the hyperbolic tangent model allows to easily marginalize over several reionization histories, it does not however predict a unique IGM temperature evolution. On the other hand, the FG and PUCH astrophysics models for reionization  allow to self-consistently account for ionization and heating  at low redshifts via both DM and astrophysical contributions.

In Sec.~\ref{sec: CMB analysis}, we have derived  new constraints on ALP decays into two photons considering the three reionization models mentioned above.  For that purpose, we have modified {\tt CLASS} to interpolate the reionization histories at low redshifts and the effective energy deposition efficiencies at higher redshifts over the DM masses and coupling to photons considered in this work. The effective energy deposition efficiencies were taken to be constant in time and equal to $f_c(x_e,z=300)$ following the findings of~\cite{Slatyer:2016qyl}. 
 We then performed an MCMC analysis including the baseline TT,TE,EE and low E Planck 2018 likelihoods and ran over a set of cosmological parameters including  the DM mass and coupling to photons as well as the reionization redshift when considering a $\tanh$ model.  Compared with previous works \cite{Cadamuro:2011fd,Bolliet:2020ofj}, our analysis improves the bounds on the DM lifetime by exploiting the full CMB anisotropy spectrum information of the latest Planck data release, and explores for the first time their dependence on the astrophysical models for reionization using the self-consistent evaluation of $x_e$. 

The summary plot of all our results is provided in Fig.~\ref{fig:reio-cons}. In particular, the three red lines delimit the regions on  the ALP parameter space (mass and coupling to photons) that are excluded at 99\%CL for the three reionization models that we have considered. 
Note however  that, by properly re-expressing the bounds on $g_{a\gamma \gamma}$ in terms of the DM lifetime, our constraints apply to any other  DM model decaying to two photons.
The limits are slightly more  stringent in the case of the PUCH model~\cite{Puchwein:2018arm}. This is expected as in the latter case the optical depth to reionization in the absence of DM energy injections is already significantly larger than the central value obtained with Planck 2018 and a $\tanh$ model.  Overall, the CMB bounds obtained here are competitive with previous existing constraints in the mass range from 20.4 eV  to 400 eV, except for  the Leo-T bound from~\cite{Wadekar:2021qae}. Let us emphasize though that the Leo-T and CMB bounds are independent as they rely on very different astrophysical and cosmological phenomena and assumptions. 
The CMB analysis has the advantage that it mostly relies on the linear evolution of the cosmological perturbations, which is well understood. However, it also partly inherits the astrophysical uncertainties on the reionization history as we show explicitly in our analysis with the three different reionization models.
In Fig.~\ref{fig:reio-cons-fut}, we have estimated the sensitivity of future CMB surveys and concluded that the latter shall give rise to constraints competitive with the Leo-T bound, even in the most conservative reionization scenarios.  
In the future, it would be interesting to go beyond the two specific reionization  models from stars considered in this work and simultaneously constrain the models by CMB, Lyman-$\alpha$ and UV and X-ray data.

Finally, in  Sec.~\ref{sec:Lycons}, we have focused on the evolution of the IGM temperature. The recent advancements in hydrodynamical simulations and in the measurements of Lyman-$\alpha$ data have allowed a determination of the IGM temperature at redshifts $z<7$. These data have been used in \cite{Liu:2020wqz} to set bounds on the decay rate to photons of DM particles with masses above 10 keV. Here we have extended this analysis to lower masses taking care of  accounting for the cooling from collisional excitations processes in the high redshift range, before the onset of reionization. Assuming a $\tanh$ model for the ionized fractions, our final constraints on the DM lifetime are illustrated in Fig.~\ref{fig:Lifetime bound} with two different reionization redshifts that correspond to the $\pm 1\sigma$ values around the central value obtained from the Planck 2018 data. The most stringent of these bounds is also reported in Fig.~\ref{fig:reio-cons} with a yellow line to ease the comparison with the CMB limits. It appears that, given  the methodology followed here, the Lyman-$\alpha$ bounds are up to one order of magnitude weaker than the CMB bounds obtained in Sec.~\ref{sec: CMB analysis}. 
Note, however, that the Lyman-$\alpha$ data has the advantage that it provides tomographic constraints,  i.e. at different redshifts, on the dark matter injection of energy whereas the CMB data is most sensitive to the integrated effect of the DM energy injection, through the parameter $\tau$ as we discussed in the main text.
It would however be interesting to review these constraints in the light of next-generation measurements of 21 cm emission/absorption that is expected to  become a very sensitive probe of the IGM temperature at low redshifts, see e.g.~\cite{Liu:2018uzy}, and potentially surpass the CMB constraint for DM decay, see also~\cite{Furlanetto:2006wp,Valdes:2007cu,Evoli:2014pva, Lopez-Honorez:2013cua}.

\vspace{0.7cm}
{\bf Note Added:}
Upon the completion of this work, we became aware of the work in \cite{Liu:2023fgu,Liu:2023nct} on a new version  of the {\tt DarkHistory} code that improves on the treatment of low-energy particles and in particular estimates the CMB constraints on the DM parameters. Their results show that the new version of the code does not affect significantly the CMB constraints derived in Sec. \ref{sec: CMB analysis} .

\section*{Acknowledgements}
We thank M. Lucca, D.C. Hooper and N. Schoeneberg for discussions  and  support on dealing with DM energy injection and reionization in {\tt CLASS} and {\tt MontePython}. We also thank T. Slatyer and H. Liu for clarifications on {\tt DarkHistory}. Furthermore, we acknowledge useful discussions with J. Torrado.
LLH is supported
by the Fonds de la Recherche Scientifique F.R.S.-FNRS through a research associate position and acknowledges support of the  FNRS research grant number F.4520.19, the ARC program of the Federation Wallonie-Bruxelles and  the IISN convention No. 4.4503.15. Computational resources have been provided by the Consortium des Equipements de Calcul Intensif (CECI), funded by the Fonds de la Recherche Scientifique de Belgique (F.R.S.-FNRS) under Grant No. 2.5020.11 and by the Walloon Region of Belgium.
RZF is supported by the Direcci\'o General de Recerca del Departament d’Empresa i Coneixement (DGR) and by the EC through the program Marie Sk\l odowska-Curie COFUND (GA 801370)-Beatriu de Pin\'os and partially by the FCT grant No.~CERN/FIS-PAR/0027/2021. 
This work has been partially supported by the MCIN/AEI/10.13039/501100011033 of Spain under grant PID2020-113644GB-I00, by the Generalitat Valenciana of Spain under the grant PROMETEO/2019/083 and by the European Union’s Framework Programme for Research and Innovation Horizon 2020 (2014–2020) under grant H2020-MSCA-ITN-2019/860881-HIDDeN.\\

\appendix

\section{Rates}
\label{sec: Rates}

In this appendix, we provide some more details on the ionization and  heating  rates contributing to the IGM temperature and ionized fractions evolutions discussed in Sec.~\ref{sec:sec2}. A more complete list of these rates can be found for example in \cite{Theuns:1998kr,Bolton:2006pc,Liu:2019bbm,Liu:2020wqz}. 

Let us first discuss the contribution to the term $\dot Y^{(0)}$ of Eq.~(\ref{eq: master equation for Tm and x}). As discussed in Sec.~\ref{sec:adiab}, there are ionization processes ($\dot x_X^\text{ion}$) 
 that increase the number of free electrons and recombination processes  ($\dot x_X^\text{rec}$) that have the opposite effect. Depending on the redshift of the process, there are two different regimes: case-B for optically thin medium, applicable at redshifts larger than the onset of reionization $z_\text{A}^\text{max}$ when the universe is mostly neutral, and case-A for optically thick medium, that applies for small redshifts $z \lesssim z_\text{A}^\text{max}$ after reionization starts.

In the optically thick regime, case-A, recombination and ionization via collisional ionization are the leading processes included in {\tt DarkHistory}. In this case, Eq.~(\ref{eq: adiabatic terms, ionization fractions}) is given by \cite{Theuns:1998kr,Bolton:2006pc,Liu:2020wqz}:
\begin{eqnarray}
	\left(\begin{matrix}  \dot{x}_\text{HII}^{(0)} \\ \dot{x}_\text{HeII}^{(0)} \\ \dot{x}_\text{HeIII}^{(0)} \end{matrix} \right) = n_e  \left(\begin{matrix}   
		 (1-x_\text{HII})  \Gamma_\text{HI}^\text{col-ion}  - x_\text{HII} \alpha^A_\text{HI}  \\
		({\cal F}_\text{He}-x_\text{HeII}-x_\text{HeIII})   \Gamma_\text{HeI}^\text{col-ion} + x_\text{HeIII} \alpha^A_\text{HeIII} - x_\text{HeII}  \left(   \Gamma_\text{HeII}^\text{col-ion} + \alpha^A_\text{HeII}\right) \\ 	 x_\text{HeII}    \Gamma_\text{HeIII}^\text{col-ion}  -x_\text{HeIII} \alpha^A_\text{HeIII} 
	\end{matrix} \right) \, .
\end{eqnarray}
where $\Gamma_\text{X}^\text{col-ion}$ are the collisional ionization rates of the different species, given e.g. in~\cite{Theuns:1998kr}, and $\subt{\alpha}{X}^A$ is the  case-A recombination  coefficient. Note that, the photoionization processes are expected to be dominated by the astrophysical sources during reionization and so are not included in $\dot Y^{(0)}$  but rather in the astro-term $\dot Y^{\rm astro}$ of Eq.~(\ref{eq: master equation for Tm and x}), see Sec.~\ref{sec:reio}.

In the optically thin regime, case-B, the processes that are instead taken into account in {\tt DarkHistory} are recombination and photoionizations given by {\tt DarkHistory}~\cite{Liu:2019bbm} 
\begin{eqnarray}	\left(\begin{matrix}  \dot{x}_\text{HII}^{(0)} \\ \dot{x}_\text{HeII}^{(0)}  \end{matrix} \right) = \left(\begin{matrix}  {\cal C}_\text{H} \left[ 4(1-x_\text{HII})\beta^B_\text{H} e^{E_{21}/T_\text{CMB}}- n_\text{H} x_e x_\text{HII} \alpha^B_\text{H}\right] \\ 
	\sum_{i=s,t} {\cal C}_{\text{He}^i} \left[ g_i ({\cal F}_\text{He}-x_\text{HeII})\beta^B_{\text{He}^i} e^{-E_{\text{He}^i}/T_\text{CMB}}- n_\text{H} x_e x_\text{HeII} \alpha^B_{\text{He}^i}  \right] 
	\end{matrix} \right)\, ,
\end{eqnarray}
where $x_e=n_e/n_H$ is the electron fraction with $n_e$ the density of free electrons and $E_{21}=10.2$ eV is the energy of the Lyman-$\alpha$ transition. Also,  $\subt{\alpha}{X}^B,\subt{\beta}{H}^B$ are  the case-B recombination and photoionization coefficients. ${\cal C}_X$ is the Peebles-C factor for the species $X$=\{Hydrogen (H), and singlet (s) and triplet (t) Helium (He$^{s,t})$\}, i.e. the probability for the species X in the $n=2$ state to decay to the ground state, and $g_i$ is the multiplicity of the state $i=s,t$ and ${\cal F}_\text{He}=n_\text{He}/n_H$.
Note that, in {\tt DarkHistory} the doubly-ionized Helium (HeIII) is only taken into account during reionization (case-A), where it is expected to be non-negligible, but not before (case-B).

Finally, the  $\dot Y^{\rm DM}$ term of Eq.~(\ref{eq: Contributions from DM injection}) accounts for the contributions from DM energy injection. The prefactor  $A$ introduced in Eq.~(\ref{eq: Contributions from DM injection})  takes the form~\cite{Liu:2019bbm}:
\begin{eqnarray}
	A =  \left(\begin{matrix} \frac{2\subt{f}{heat}}{3\left(1+\subt{{\cal F}}{He} +x_e\right)  }  \\ 
		\frac{1}{\subt{{\cal R}}{HI} }  \left( \subt{f}{HII} + \frac{4}{3} (1-{\cal C}_H) \subt{f}{exc}  \right) \\  \frac{\subt{f}{HeII}}{\subt{{\cal R}}{HeI} }  
		\\ \frac{\subt{f}{HeIII}}{\subt{{\cal R}}{HeII} }    \end{matrix} \right)
\end{eqnarray}
where $\subt{{\cal R}}{X}$ is the ionization potential of the atom/ion X.

\section{Modified {\tt DarkHistory}}
\label{sec:modif_darkhistory}

 While running {\tt DarkHistory} for sub-keV masses we observed a sudden drop of the matter temperature at redshifts near the beginning of reionization. We found the effect to be more prominent for DM masses around 100 eV and couplings $g_{a\gamma \gamma}$ larger than a few times $10^{-13}$ GeV$^{-1}$. We noticed that this drop is much less significant when considering  larger masses for the decaying DM.

\begin{figure}
	\centering
	\includegraphics[width=0.46\linewidth]{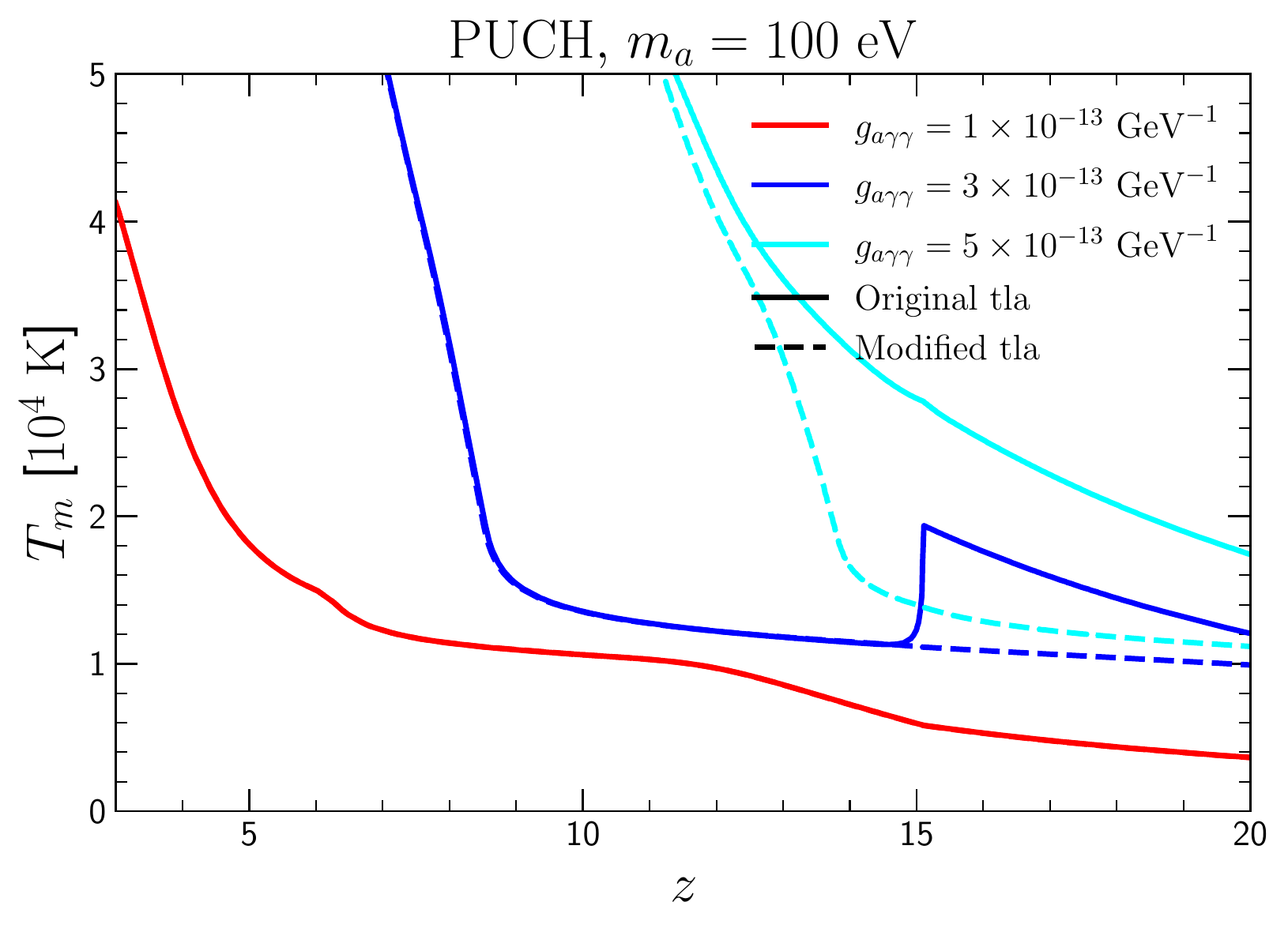}
	\includegraphics[width=0.46\linewidth]{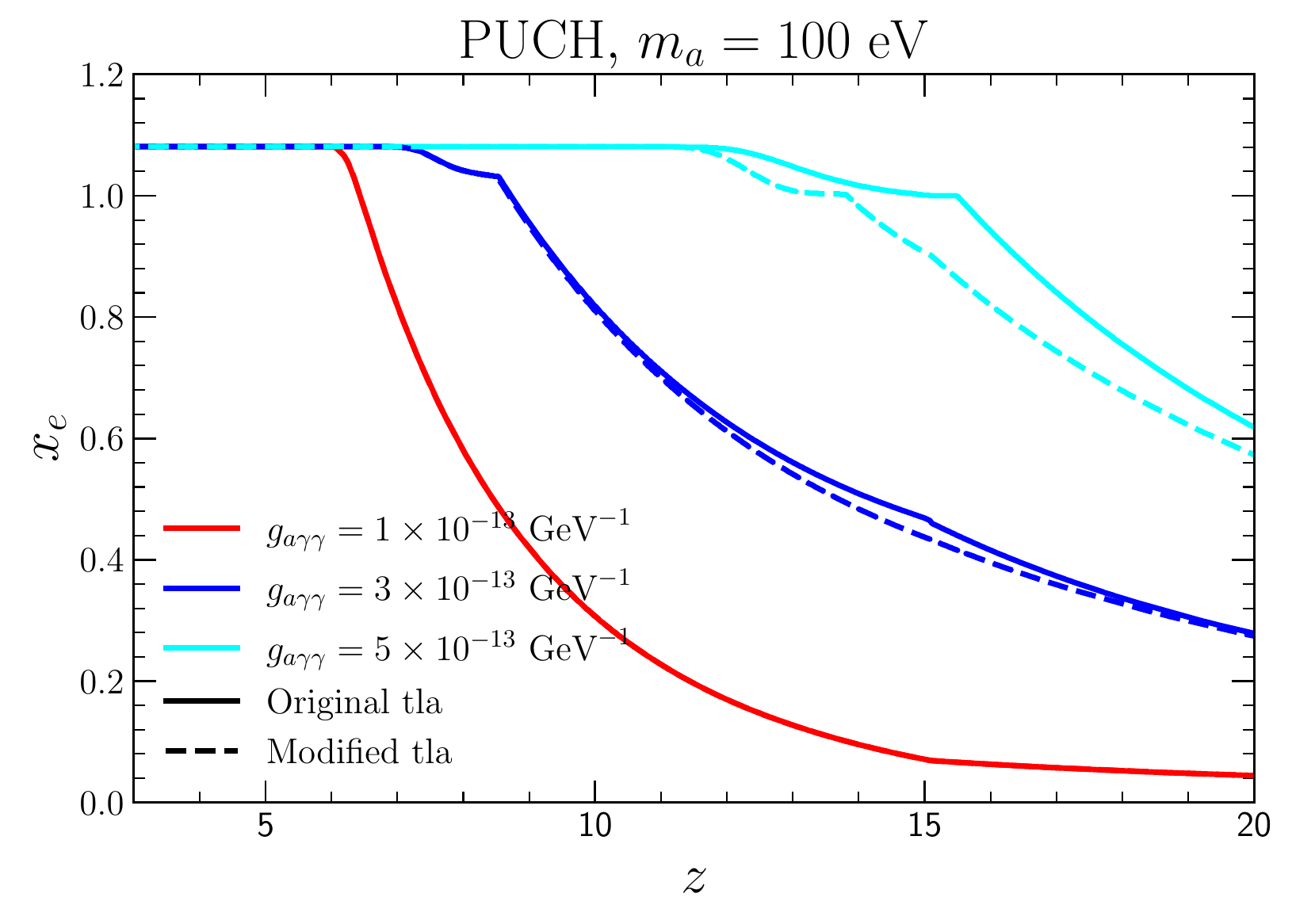}
 \includegraphics[width=0.46\linewidth]{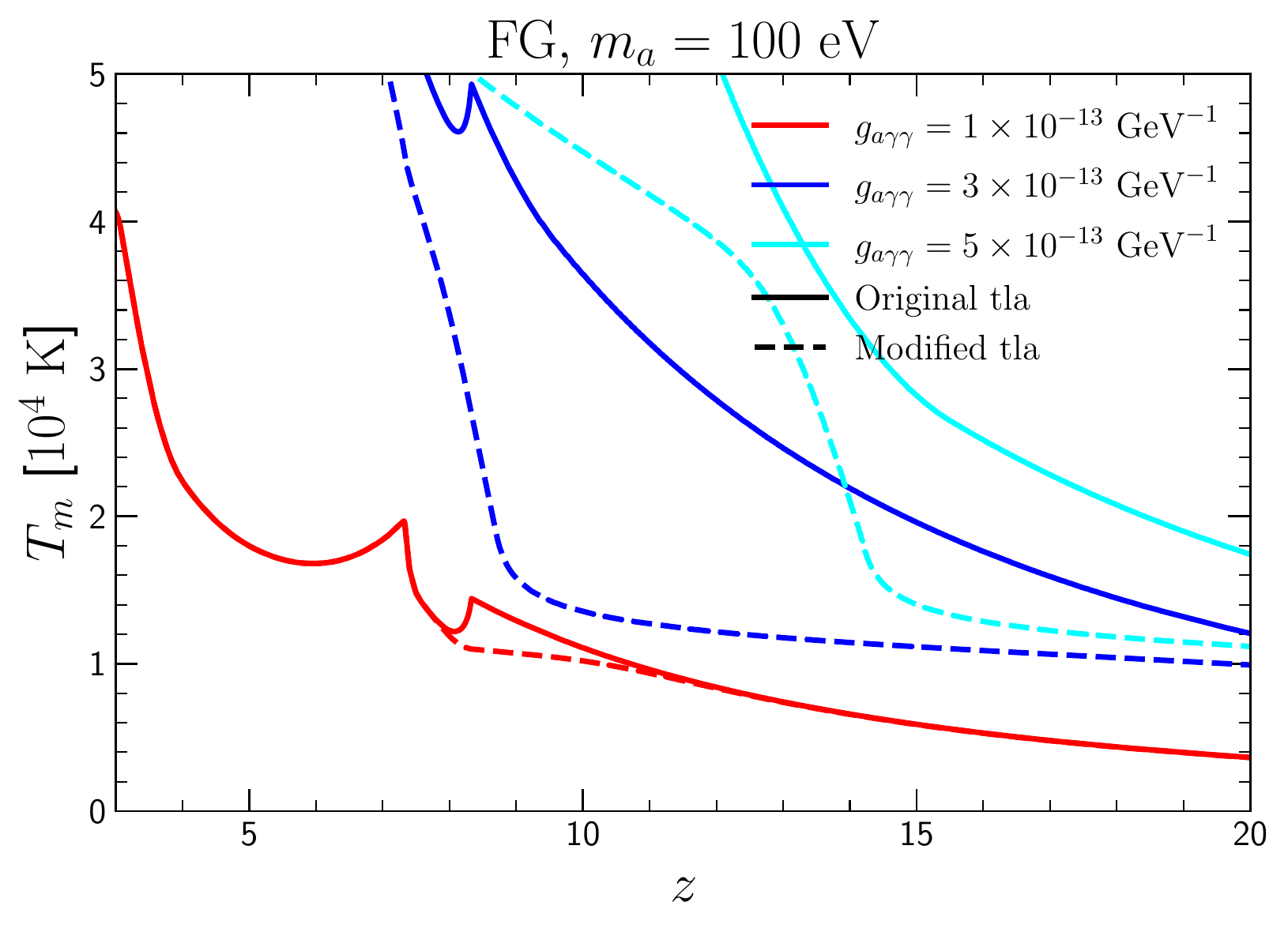}
	\includegraphics[width=0.46\linewidth]{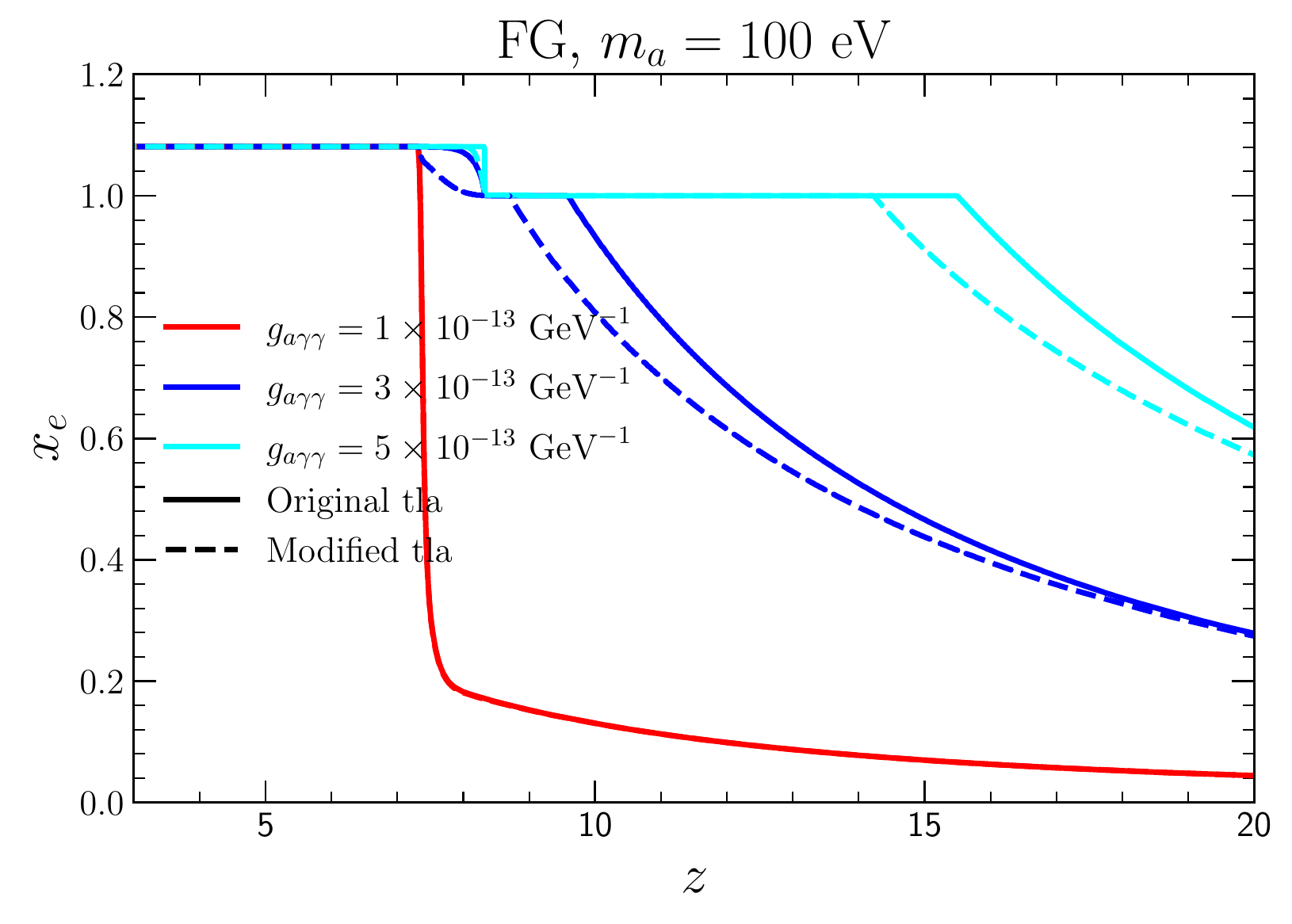}
	\caption{Comparison between the reionization history obtained using the original version of {\tt DarkHistory} (continuous lines) and our modified version (dashed lines), in terms of the matter temperature $T_m$ (left panel) and ionization fraction $x_e$ (right panel). The reionization history in the top panels is obtained using PUCH, whereas FG is employed for the bottom panels.}
	\label{fig:tm_xe_modified_darkhistory}
\end{figure}

 It appears that the origin of this sudden drop is related to the (non-)inclusion of collisional excitations at redshifts around the onset of reionization, that we denote with $z_A^\text{max}$, in both the {\tt TIGM} and {\tt Master} branches of {\tt DarkHistory} used here.  These processes have the largest cooling rates at the threshold of reionization (at least for the DM masses and couplings mentioned above) and they are included for $z<z_{A}^{\rm max}$ (in the reionization part of the code) by default, with e.g.~$z_A^{\rm max}=15.1$ for the PUCH model. If we do not include such processes for $z>z_A^{\rm max}$,  we get a sudden drop in $T_{m}$ at the matching point. 
We have modified the \textit{tla.py} file, which includes the evolution of the matter temperature so that the collisional excitation terms are also included before the matching. We have done so in both the {\tt Master} (used in Secs~\ref{sec:reio} and~\ref{sec: CMB analysis}) and {\tt TIGM} (used in Sec.~\ref{sec:Lycons}) branches of the {\tt DarkHistory v1.1} code. Using this modified version, the sudden drop in the matter temperature disappeared without changing the pre-reionization evolution.  The temperature drop can also have, indirectly, a significant impact on the ionization history when 
dark matter energy injection is relatively large. However, we find that the  $x_e$ evolution is very weakly affected for ALP masses and couplings at the limit of our exclusion bounds obtained in our CMB analysis, see Sec.~\ref{sec:CMBcons}.
This can be seen in Fig.~\ref{fig:tm_xe_modified_darkhistory}, which shows a comparison between the reionization history obtained using the original version of the code (continuous lines) and our modified version (dashed lines), assuming $m_a=100$ eV. In the top panels we use the PUCH model with $z_A^{\rm max}=15.1$, whereas in the bottom panels, the FG rates are employed with $z_A^{\rm max}=7.8$. The sudden drop obtained with the original code around $z_A^{\rm max}$ disappears with our modifications. We have also tested how the inclusion of additional cooling/heating terms in the pre-reionization evolution, such as collisional ionization and bremsstrahlung, would affect the results but we did not find sizeable effects, which can be understood from the fact that these rates are subdominant.

\
\bibliography{biblio}

\providecommand{\href}[2]{#2}\begingroup\raggedright\begin{thebibliography}{10}

\bibitem{Planck:2018vyg}
{\scshape Planck} collaboration, \emph{{Planck 2018 results. VI. Cosmological
  parameters}},
  \href{https://doi.org/10.1051/0004-6361/201833910}{\emph{Astron. Astrophys.}
  {\bfseries 641} (2020) A6}
  [\href{https://arxiv.org/abs/1807.06209}{{\ttfamily 1807.06209}}].

\bibitem{Walther:2018pnn}
M.~Walther, J.~O\~norbe, J.F.~Hennawi and Z.~Luki\'c, \emph{{New Constraints on
  IGM Thermal Evolution from the Ly\ensuremath{\alpha} Forest Power Spectrum}},
  \href{https://doi.org/10.3847/1538-4357/aafad1}{\emph{Astrophys. J.}
  {\bfseries 872} (2019) 13}
  [\href{https://arxiv.org/abs/1808.04367}{{\ttfamily 1808.04367}}].

\bibitem{Gaikwad:2020art}
P.~Gaikwad et~al., \emph{{Probing the thermal state of the intergalactic medium
  at z \ensuremath{>} 5 with the transmission spikes in high-resolution Ly
  \ensuremath{\alpha} forest spectra}},
  \href{https://doi.org/10.1093/mnras/staa907}{\emph{Mon. Not. Roy. Astron.
  Soc.} {\bfseries 494} (2020) 5091}
  [\href{https://arxiv.org/abs/2001.10018}{{\ttfamily 2001.10018}}].

\bibitem{Takahashi:2021tff}
F.~Takahashi and W.~Yin, \emph{{Challenges for heavy QCD axion inflation}},
  \href{https://doi.org/10.1088/1475-7516/2021/10/057}{\emph{JCAP} {\bfseries
  10} (2021) 057} [\href{https://arxiv.org/abs/2105.10493}{{\ttfamily
  2105.10493}}].

\bibitem{Gelmini:2021yzu}
G.B.~Gelmini, A.~Simpson and E.~Vitagliano, \emph{{Gravitational waves from
  axionlike particle cosmic string-wall networks}},
  \href{https://doi.org/10.1103/PhysRevD.104.L061301}{\emph{Phys. Rev. D}
  {\bfseries 104} (2021) 061301}
  [\href{https://arxiv.org/abs/2103.07625}{{\ttfamily 2103.07625}}].

\bibitem{Bernal:2022xyi}
J.L.~Bernal, A.~Caputo, G.~Sato-Polito, J.~Mirocha and M.~Kamionkowski,
  \emph{{Seeking dark matter with $\gamma$-ray attenuation}},
  \href{https://arxiv.org/abs/2208.13794}{{\ttfamily 2208.13794}}.

\bibitem{Branco:2023frw}
N.P.~Branco, R.Z.~Ferreira and J.a.G.~Rosa, \emph{{Superradiant axion clouds
  around asteroid-mass primordial black holes}},
  \href{https://arxiv.org/abs/2301.01780}{{\ttfamily 2301.01780}}.

\bibitem{Carenza:2023qxh}
P.~Carenza, G.~Lucente and E.~Vitagliano, \emph{{Probing the Blue Axion with
  Cosmic Optical Background Anisotropies}},
  \href{https://arxiv.org/abs/2301.06560}{{\ttfamily 2301.06560}}.

\bibitem{Slatyer:2015jla}
T.R.~Slatyer, \emph{{Indirect dark matter signatures in the cosmic dark ages.
  I. Generalizing the bound on s-wave dark matter annihilation from Planck
  results}}, \href{https://doi.org/10.1103/PhysRevD.93.023527}{\emph{Phys. Rev.
  D} {\bfseries 93} (2016) 023527}
  [\href{https://arxiv.org/abs/1506.03811}{{\ttfamily 1506.03811}}].

\bibitem{Slatyer:2015kla}
T.R.~Slatyer, \emph{{Indirect Dark Matter Signatures in the Cosmic Dark Ages
  II. Ionization, Heating and Photon Production from Arbitrary Energy
  Injections}}, \href{https://doi.org/10.1103/PhysRevD.93.023521}{\emph{Phys.
  Rev. D} {\bfseries 93} (2016) 023521}
  [\href{https://arxiv.org/abs/1506.03812}{{\ttfamily 1506.03812}}].

\bibitem{Cadamuro:2011fd}
D.~Cadamuro and J.~Redondo, \emph{{Cosmological bounds on pseudo
  Nambu-Goldstone bosons}},
  \href{https://doi.org/10.1088/1475-7516/2012/02/032}{\emph{JCAP} {\bfseries
  02} (2012) 032} [\href{https://arxiv.org/abs/1110.2895}{{\ttfamily
  1110.2895}}].

\bibitem{Bolliet:2020ofj}
B.~Bolliet, J.~Chluba and R.~Battye, \emph{{Spectral distortion constraints on
  photon injection from low-mass decaying particles}},
  \href{https://doi.org/10.1093/mnras/stab1997}{\emph{Mon. Not. Roy. Astron.
  Soc.} {\bfseries 507} (2021) 3148}
  [\href{https://arxiv.org/abs/2012.07292}{{\ttfamily 2012.07292}}].

\bibitem{Liu:2019bbm}
H.~Liu, G.W.~Ridgway and T.R.~Slatyer, \emph{{Code package for calculating
  modified cosmic ionization and thermal histories with dark matter and other
  exotic energy injections}},
  \href{https://doi.org/10.1103/PhysRevD.101.023530}{\emph{Phys. Rev. D}
  {\bfseries 101} (2020) 023530}
  [\href{https://arxiv.org/abs/1904.09296}{{\ttfamily 1904.09296}}].

\bibitem{Faucher-Giguere:2019kbp}
C.-A.~Faucher-Gigu\`ere, \emph{{A cosmic UV/X-ray background model update}},
  \href{https://doi.org/10.1093/mnras/staa302}{\emph{Mon. Not. Roy. Astron.
  Soc.} {\bfseries 493} (2020) 1614}
  [\href{https://arxiv.org/abs/1903.08657}{{\ttfamily 1903.08657}}].

\bibitem{Puchwein:2018arm}
E.~Puchwein, F.~Haardt, M.G.~Haehnelt and P.~Madau, \emph{{Consistent modelling
  of the meta-galactic UV background and the thermal/ionization history of the
  intergalactic medium}},
  \href{https://doi.org/10.1093/mnras/stz222}{\emph{Mon. Not. Roy. Astron.
  Soc.} {\bfseries 485} (2019) 47}
  [\href{https://arxiv.org/abs/1801.04931}{{\ttfamily 1801.04931}}].

\bibitem{Wadekar:2021qae}
D.~Wadekar and Z.~Wang, \emph{{Strong constraints on decay and annihilation of
  dark matter from heating of gas-rich dwarf galaxies}},
  \href{https://doi.org/10.1103/PhysRevD.106.075007}{\emph{Phys. Rev. D}
  {\bfseries 106} (2022) 075007}
  [\href{https://arxiv.org/abs/2111.08025}{{\ttfamily 2111.08025}}].

\bibitem{Liu:2020wqz}
H.~Liu, W.~Qin, G.W.~Ridgway and T.R.~Slatyer, \emph{{Lyman-\ensuremath{\alpha}
  constraints on cosmic heating from dark matter annihilation and decay}},
  \href{https://doi.org/10.1103/PhysRevD.104.043514}{\emph{Phys. Rev. D}
  {\bfseries 104} (2021) 043514}
  [\href{https://arxiv.org/abs/2008.01084}{{\ttfamily 2008.01084}}].

\bibitem{Peebles:1968ja}
P.J.E.~Peebles, \emph{{Recombination of the Primeval Plasma}},
  \href{https://doi.org/10.1086/149628}{\emph{Astrophys. J.} {\bfseries 153}
  (1968) 1}.

\bibitem{Zeldovich:1969ff}
Y.B.~Zeldovich and R.A.~Sunyaev, \emph{{The Interaction of Matter and Radiation
  in a Hot-Model Universe}},
  \href{https://doi.org/10.1007/BF00661821}{\emph{Astrophys. Space Sci.}
  {\bfseries 4} (1969) 301}.

\bibitem{Theuns:1998kr}
T.~Theuns, A.~Leonard, G.~Efstathiou, F.R.~Pearce and P.A.~Thomas,
  \emph{{P**3M-SPH simulations of the lyman-alpha forest}},
  \href{https://doi.org/10.1046/j.1365-8711.1998.02040.x}{\emph{Mon. Not. Roy.
  Astron. Soc.} {\bfseries 301} (1998) 478}
  [\href{https://arxiv.org/abs/astro-ph/9805119}{{\ttfamily
  astro-ph/9805119}}].

\bibitem{Bolton:2006pc}
J.S.~Bolton and M.G.~Haehnelt, \emph{{The nature and evolution of the highly
  ionized near-zones in the absorption spectra of z\textasciitilde{}=6
  quasars}}, \href{https://doi.org/10.1111/j.1365-2966.2006.11176.x}{\emph{Mon.
  Not. Roy. Astron. Soc.} {\bfseries 374} (2007) 493}
  [\href{https://arxiv.org/abs/astro-ph/0607331}{{\ttfamily
  astro-ph/0607331}}].

\bibitem{Slatyer:2009yq}
T.R.~Slatyer, N.~Padmanabhan and D.P.~Finkbeiner, \emph{{CMB Constraints on
  WIMP Annihilation: Energy Absorption During the Recombination Epoch}},
  \href{https://doi.org/10.1103/PhysRevD.80.043526}{\emph{Phys. Rev. D}
  {\bfseries 80} (2009) 043526}
  [\href{https://arxiv.org/abs/0906.1197}{{\ttfamily 0906.1197}}].

\bibitem{Lopez-Honorez:2013cua}
L.~Lopez-Honorez, O.~Mena, S.~Palomares-Ruiz and A.C.~Vincent,
  \emph{{Constraints on dark matter annihilation from CMB observationsbefore
  Planck}}, \href{https://doi.org/10.1088/1475-7516/2013/07/046}{\emph{JCAP}
  {\bfseries 07} (2013) 046} [\href{https://arxiv.org/abs/1303.5094}{{\ttfamily
  1303.5094}}].

\bibitem{Poulin:2015pna}
V.~Poulin, P.D.~Serpico and J.~Lesgourgues, \emph{{Dark Matter annihilations in
  halos and high-redshift sources of reionization of the universe}},
  \href{https://doi.org/10.1088/1475-7516/2015/12/041}{\emph{JCAP} {\bfseries
  12} (2015) 041} [\href{https://arxiv.org/abs/1508.01370}{{\ttfamily
  1508.01370}}].

\bibitem{Lesgourgues:2011re}
J.~Lesgourgues, \emph{{The Cosmic Linear Anisotropy Solving System (CLASS) I:
  Overview}},  \href{https://arxiv.org/abs/1104.2932}{{\ttfamily 1104.2932}}.

\bibitem{Blas:2011rf}
D.~Blas, J.~Lesgourgues and T.~Tram, \emph{{The Cosmic Linear Anisotropy
  Solving System (CLASS) II: Approximation schemes}},
  \href{https://doi.org/10.1088/1475-7516/2011/07/034}{\emph{JCAP} {\bfseries
  07} (2011) 034} [\href{https://arxiv.org/abs/1104.2933}{{\ttfamily
  1104.2933}}].

\bibitem{Lesgourgues:2013bra}
J.~Lesgourgues and T.~Tram, \emph{{Fast and accurate CMB computations in
  non-flat FLRW universes}},
  \href{https://doi.org/10.1088/1475-7516/2014/09/032}{\emph{JCAP} {\bfseries
  09} (2014) 032} [\href{https://arxiv.org/abs/1312.2697}{{\ttfamily
  1312.2697}}].

\bibitem{Lewis:2008wr}
A.~Lewis, \emph{{Cosmological parameters from WMAP 5-year temperature maps}},
  \href{https://doi.org/10.1103/PhysRevD.78.023002}{\emph{Phys. Rev. D}
  {\bfseries 78} (2008) 023002}
  [\href{https://arxiv.org/abs/0804.3865}{{\ttfamily 0804.3865}}].

\bibitem{Shull:1985}
J.M.~Shull and M.E.~van Steenberg, \emph{{X-ray secondary heating and
  ionization in quasar emission-line clouds}},
  \href{https://doi.org/10.1086/163605}{\emph{Astrophys. J.} {\bfseries 298}
  (1985) 268}.

\bibitem{Chen:2003gz}
X.-L.~Chen and M.~Kamionkowski, \emph{{Particle decays during the cosmic dark
  ages}}, \href{https://doi.org/10.1103/PhysRevD.70.043502}{\emph{Phys. Rev.}
  {\bfseries D70} (2004) 043502}
  [\href{https://arxiv.org/abs/astro-ph/0310473}{{\ttfamily
  astro-ph/0310473}}].

\bibitem{Padmanabhan:2005es}
N.~Padmanabhan and D.P.~Finkbeiner, \emph{{Detecting dark matter annihilation
  with CMB polarization: Signatures and experimental prospects}},
  \href{https://doi.org/10.1103/PhysRevD.72.023508}{\emph{Phys. Rev. D}
  {\bfseries 72} (2005) 023508}
  [\href{https://arxiv.org/abs/astro-ph/0503486}{{\ttfamily
  astro-ph/0503486}}].

\bibitem{Diamanti:2013bia}
R.~Diamanti, L.~Lopez-Honorez, O.~Mena, S.~Palomares-Ruiz and A.C.~Vincent,
  \emph{{Constraining Dark Matter Late-Time Energy Injection: Decays and P-Wave
  Annihilations}},
  \href{https://doi.org/10.1088/1475-7516/2014/02/017}{\emph{JCAP} {\bfseries
  02} (2014) 017} [\href{https://arxiv.org/abs/1308.2578}{{\ttfamily
  1308.2578}}].

\bibitem{Slatyer:2016qyl}
T.R.~Slatyer and C.-L.~Wu, \emph{{General Constraints on Dark Matter Decay from
  the Cosmic Microwave Background}},
  \href{https://doi.org/10.1103/PhysRevD.95.023010}{\emph{Phys. Rev. D}
  {\bfseries 95} (2017) 023010}
  [\href{https://arxiv.org/abs/1610.06933}{{\ttfamily 1610.06933}}].

\bibitem{CLASS}
J.~Lesgourgues, \emph{The cosmic linear anisotropy solving system (class) i:
  Overview},  2011.
\newblock 10.48550/ARXIV.1104.2932.

\bibitem{Stocker:2018avm}
P.~St\"ocker, M.~Kr\"amer, J.~Lesgourgues and V.~Poulin, \emph{{Exotic energy
  injection with ExoCLASS: Application to the Higgs portal model and
  evaporating black holes}},
  \href{https://doi.org/10.1088/1475-7516/2018/03/018}{\emph{JCAP} {\bfseries
  03} (2018) 018} [\href{https://arxiv.org/abs/1801.01871}{{\ttfamily
  1801.01871}}].

\bibitem{Lucca:2019rxf}
M.~Lucca, N.~Sch\"oneberg, D.C.~Hooper, J.~Lesgourgues and J.~Chluba,
  \emph{{The synergy between CMB spectral distortions and anisotropies}},
  \href{https://doi.org/10.1088/1475-7516/2020/02/026}{\emph{JCAP} {\bfseries
  02} (2020) 026} [\href{https://arxiv.org/abs/1910.04619}{{\ttfamily
  1910.04619}}].

\bibitem{Ali-Haimoud:2010tlj}
Y.~Ali-Haimoud and C.M.~Hirata, \emph{{Ultrafast effective multi-level atom
  method for primordial hydrogen recombination}},
  \href{https://doi.org/10.1103/PhysRevD.82.063521}{\emph{Phys. Rev. D}
  {\bfseries 82} (2010) 063521}
  [\href{https://arxiv.org/abs/1006.1355}{{\ttfamily 1006.1355}}].

\bibitem{Lee:2020obi}
N.~Lee and Y.~Ali-Ha\"\i{}moud, \emph{{HYREC-2: a highly accurate
  sub-millisecond recombination code}},
  \href{https://doi.org/10.1103/PhysRevD.102.083517}{\emph{Phys. Rev. D}
  {\bfseries 102} (2020) 083517}
  [\href{https://arxiv.org/abs/2007.14114}{{\ttfamily 2007.14114}}].

\bibitem{Finkbeiner:2011dx}
D.P.~Finkbeiner, S.~Galli, T.~Lin and T.R.~Slatyer, \emph{{Searching for Dark
  Matter in the CMB: A Compact Parameterization of Energy Injection from New
  Physics}}, \href{https://doi.org/10.1103/PhysRevD.85.043522}{\emph{Phys. Rev.
  D} {\bfseries 85} (2012) 043522}
  [\href{https://arxiv.org/abs/1109.6322}{{\ttfamily 1109.6322}}].

\bibitem{Brinckmann:2018cvx}
T.~Brinckmann and J.~Lesgourgues, \emph{{MontePython 3: boosted MCMC sampler
  and other features}},
  \href{https://doi.org/10.1016/j.dark.2018.100260}{\emph{Phys. Dark Univ.}
  {\bfseries 24} (2019) 100260}
  [\href{https://arxiv.org/abs/1804.07261}{{\ttfamily 1804.07261}}].

\bibitem{Abazajian:2019eic}
K.~Abazajian et~al., \emph{{CMB-S4 Science Case, Reference Design, and Project
  Plan}},  \href{https://arxiv.org/abs/1907.04473}{{\ttfamily 1907.04473}}.

\bibitem{Watts:2018etg}
D.J.~Watts et~al., \emph{{A Projected Estimate of the Reionization Optical
  Depth Using the CLASS Experiment\textquoteright{}s Sample Variance Limited
  E-mode Measurement}},
  \href{https://doi.org/10.3847/1538-4357/aad283}{\emph{Astrophys. J.}
  {\bfseries 863} (2018) 121}
  [\href{https://arxiv.org/abs/1801.01481}{{\ttfamily 1801.01481}}].

\bibitem{Becker:2010cu}
G.D.~Becker, J.S.~Bolton, M.G.~Haehnelt and W.L.W.~Sargent, \emph{{Detection of
  Extended He II Reionization in the Temperature Evolution of the Intergalactic
  Medium}}, \href{https://doi.org/10.1111/j.1365-2966.2010.17507.x}{\emph{Mon.
  Not. Roy. Astron. Soc.} {\bfseries 410} (2011) 1096}
  [\href{https://arxiv.org/abs/1008.2622}{{\ttfamily 1008.2622}}].

\bibitem{Liu:2018uzy}
H.~Liu and T.R.~Slatyer, \emph{{Implications of a 21-cm signal for dark matter
  annihilation and decay}},
  \href{https://doi.org/10.1103/PhysRevD.98.023501}{\emph{Phys. Rev. D}
  {\bfseries 98} (2018) 023501}
  [\href{https://arxiv.org/abs/1803.09739}{{\ttfamily 1803.09739}}].

\bibitem{Furlanetto:2006wp}
S.R.~Furlanetto, S.P.~Oh and E.~Pierpaoli, \emph{{The Effects of Dark Matter
  Decay and Annihilation on the High-Redshift 21 cm Background}},
  \href{https://doi.org/10.1103/PhysRevD.74.103502}{\emph{Phys. Rev. D}
  {\bfseries 74} (2006) 103502}
  [\href{https://arxiv.org/abs/astro-ph/0608385}{{\ttfamily
  astro-ph/0608385}}].

\bibitem{Valdes:2007cu}
M.~Valdes, A.~Ferrara, M.~Mapelli and E.~Ripamonti, \emph{{Constraining DM
  through 21 cm observations}},
  \href{https://doi.org/10.1111/j.1365-2966.2007.11594.x}{\emph{Mon. Not. Roy.
  Astron. Soc.} {\bfseries 377} (2007) 245}
  [\href{https://arxiv.org/abs/astro-ph/0701301}{{\ttfamily
  astro-ph/0701301}}].

\bibitem{Evoli:2014pva}
C.~Evoli, A.~Mesinger and A.~Ferrara, \emph{{Unveiling the nature of dark
  matter with high redshift 21 cm line experiments}},
  \href{https://doi.org/10.1088/1475-7516/2014/11/024}{\emph{JCAP} {\bfseries
  11} (2014) 024} [\href{https://arxiv.org/abs/1408.1109}{{\ttfamily
  1408.1109}}].

\bibitem{Liu:2023fgu}
H.~Liu, W.~Qin, G.W.~Ridgway and T.R.~Slatyer, \emph{{Exotic energy injection
  in the early universe I: a novel treatment for low-energy electrons and
  photons}},  \href{https://arxiv.org/abs/2303.07366}{{\ttfamily 2303.07366}}.

\bibitem{Liu:2023nct}
H.~Liu, W.~Qin, G.W.~Ridgway and T.R.~Slatyer, \emph{{Exotic energy injection
  in the early universe II: CMB spectral distortions and constraints on light
  dark matter}},  \href{https://arxiv.org/abs/2303.07370}{{\ttfamily
  2303.07370}}.

\end{thebibliography}\endgroup
\end{document}